\documentclass[ twocolumn, superscriptaddress, amsmath,amssymb, prc, aps ]{revtex4-1}

\RequirePackage{fix-cm}
\RequirePackage{graphicx}
 \RequirePackage{mathptmx}      % use Times fonts if available on your TeX system

\usepackage[%  
    colorlinks=true,
    linkcolor=red
]{hyperref}

\usepackage{multirow}
\begin{document}

\title{Development of ultra-pure NaI(Tl) detectors for the COSINE-200 experiment%\thanksref{t1}
}

\author{B.J.~Park}
\affiliation{IBS School, University of Science and Technology (UST), Daejeon 34113, Republic of Korea}
\affiliation{Center for Underground Physics, Institute for Basic Science (IBS), Daejeon 34126, Republic of Korea}
\author{J.J.~Choi}
\affiliation{Department of Physics and Astronomy, Seoul National University, Seoul 08826, Republic of Korea}
\author{J.S.~Choe}
\affiliation{Center for Underground Physics, Institute for Basic Science (IBS), Daejeon 34126, Republic of Korea}
\author{O.~Gileva}
\affiliation{Center for Underground Physics, Institute for Basic Science (IBS), Daejeon 34126, Republic of Korea}
\author{C.~Ha}
\email{changhyon.ha@gmail.com}
\affiliation{Department of Physics, Chung-Ang University, Seoul 06973, Republic of Korea}
\author{A.~Iltis}
\affiliation{Damavan Imaging, Troyes 10430, France}
\author{E.J.~Jeon}
\affiliation{Center for Underground Physics, Institute for Basic Science (IBS), Daejeon 34126, Republic of Korea}
\author{D.Y.~Kim}
\affiliation{Center for Underground Physics, Institute for Basic Science (IBS), Daejeon 34126, Republic of Korea}
\author{K.W.~Kim}
\affiliation{Center for Underground Physics, Institute for Basic Science (IBS), Daejeon 34126, Republic of Korea}
\author{S.K.~Kim}
\affiliation{Department of Physics and Astronomy, Seoul National University, Seoul 08826, Republic of Korea}
\author{Y.D.~Kim}
\affiliation{Center for Underground Physics, Institute for Basic Science (IBS), Daejeon 34126, Republic of Korea}
\affiliation{IBS School, University of Science and Technology (UST), Daejeon 34113, Republic of Korea}
\author{Y.J.~Ko}
\affiliation{Center for Underground Physics, Institute for Basic Science (IBS), Daejeon 34126, Republic of Korea}
\author{C.H.~Lee}
\affiliation{Center for Underground Physics, Institute for Basic Science (IBS), Daejeon 34126, Republic of Korea}
\author{H.S.~Lee}
\email{hyunsulee@ibs.re.kr}
\affiliation{Center for Underground Physics, Institute for Basic Science (IBS), Daejeon 34126, Republic of Korea}
\affiliation{IBS School, University of Science and Technology (UST), Daejeon 34113, Republic of Korea}
\author{I.S.~Lee}
\affiliation{Center for Underground Physics, Institute for Basic Science (IBS), Daejeon 34126, Republic of Korea}
\author{M.H.~Lee}
\affiliation{Center for Underground Physics, Institute for Basic Science (IBS), Daejeon 34126, Republic of Korea}
\affiliation{IBS School, University of Science and Technology (UST), Daejeon 34113, Republic of Korea}
\author{S.H.~Lee}
\affiliation{IBS School, University of Science and Technology (UST), Daejeon 34113, Republic of Korea}
\affiliation{Center for Underground Physics, Institute for Basic Science (IBS), Daejeon 34126, Republic of Korea}
\author{S.J.~Ra}
\affiliation{Center for Underground Physics, Institute for Basic Science (IBS), Daejeon 34126, Republic of Korea}
\author{J.K.~Son}
\affiliation{Center for Underground Physics, Institute for Basic Science (IBS), Daejeon 34126, Republic of Korea}
\author{K.A.~Shin}
\affiliation{Center for Underground Physics, Institute for Basic Science (IBS), Daejeon 34126, Republic of Korea}

\begin{abstract}
The annual modulation signal observed by the DAMA experiment is a long-standing question in the community of dark matter direct detection. 
This necessitates an independent verification of its existence using the same detection technique. 
The COSINE-100 experiment has been operating with 106~kg of low-background NaI(Tl) detectors providing interesting checks on the DAMA signal.  
However, due to higher backgrounds in the NaI(Tl) crystals used in COSINE-100 relative to those used for DAMA, it was difficult to reach final conclusions. 
Since the start of COSINE-100 data taking in 2016, we also have initiated a program to develop ultra-pure NaI(Tl) crystals for COSINE-200, the next phase of the experiment. 
The program includes efforts of raw powder purification, ultra-pure NaI(Tl) crystal growth, and detector assembly techniques. 
After extensive research and development of NaI(Tl) crystal growth, we have successfully grown a few small-size (0.61$-$0.78 kg) thallium-doped crystals with high radio-purity.
A high light yield has been achieved by improvements of our detector assembly technique.
Here we report the ultra-pure NaI(Tl) detector developments at the Institute for Basic Science, Korea.
The technique developed here will be applied to the production of NaI(Tl) detectors for the COSINE-200 experiment. 

\end{abstract}

\maketitle

\section{Introduction}
\label{intro}
Although numerous astronomical observations support the conclusion that most of the matter in the universe is invisible dark matter, an understanding of its nature and interactions remains elusive~\cite{Clowe:2006eq,Aghanim:2018eyx,Bertone:2016nfn}. 
Dark matter phenomenon can be explained by new particles, such as weakly interacting massive particles~(WIMPs)~\cite{PhysRevLett.39.165,Goodman:1984dc}.
Even though tremendous efforts to search for evidence of WIMP dark matter by directly detecting nuclei recoiling from WIMP-nucleus interactions are being pursued, no definitive signal has been observed~\cite{Undagoitia:2015gya,Schumann:2019eaa}. One exception is the DAMA experiment that, using an array of NaI(Tl) detectors~\cite{Bernabei:2013xsa,Bernabei:2018yyw}, observes an annual modulation of event rates that can be interpreted as resulting from WIMP-nuclei interactions~\cite{Savage:2008er,Ko:2019enb}. 
However, this result has been the subject of a continuing debate because the WIMP-nucleon cross sections inferred from the DAMA modulation are in conflict with limits from other experiments~\cite{Tanabashi:2018oca,Kim:2012rza,Angloher:2014myn,Agnese:2014aze,Abe:2015eos,PhysRevLett.118.021303,Aprile:2017yea,Agnese:2017njq,Aprile:2018dbl,Agnes:2018ves,Akerib:2018zoq,Abdelhameed:2019hmk}. 

An unambiguous verification of the DAMA signal by independent experiments using the same NaI(Tl) crystal
target material is mandatory. Experimental efforts by several groups using NaI(Tl) as the target medium are
currently underway~\cite{Kim:2014toa,sabre,Adhikari:2017esn,Fushimi:2018qzk,Coarasa:2018qzs,Amare:2018sxx,Suerfu:2019snq}. 
Even though interesting physics results from COSINE-100~\cite{Adhikari:2018ljm,Adhikari:2019off} and ANAIS-112~\cite{Amare:2019jul} have been reported, 
none of these efforts have yet achieved a background level similar or lower than that of the DAMA experiment. 
This large background is a strong impediment to the establishment of any unambiguous conclusion of the DAMA's observation. 

From our COSINE-100 experience, we have identified the crystal growth as the key component
for reducing the overall background, of which radioactive $^{40}$K and $^{210}$Pb contaminants
are the most troublesome~\cite{Amare:2018sxx,adhikari16,cosinebg}. 
The COSINE-100 detector is immersed in an active liquid scintillator veto system that
reduces the impact of $^{40}$K impurities~\cite{Adhikari:2017esn,Park:2017jvs,Adhikari:2020asl}
but $^{210}$Pb has to be eliminated at the crystal production stage.

The PICO-LON experiment made a crystal with a $^{210}$Pb contamination as low as 29$\pm$7~$\mu$Bq/kg by using a cation resin-based NaI powder purification system, but had difficulties reproducing additional crystals with similar quality~\cite{Fushimi:2018qzk,Kanemitsu_2020}. The SABRE experiment recently produced a NaI(Tl) crystal with $^{210}$Pb and $^{40}$K contaminations of 0.34 $\pm$ 0.04 mBq/kg and 4.3 $\pm$ 0.2 ppb~\cite{Suerfu:2019snq}, respectively. 

As part of an effort to upgrade the on-going COSINE-100 experiment to the next-phase COSINE-200 experiment,
we have conducted a R\&D aimed at low-background NaI(Tl) crystal growth.
An initial chemical purification stage based on a recrystallization method demonstrated sufficient reduction of K and Pb in the powder~\cite{Shin:2018ioq}.
Based on this, a mass production facility for the powder purification was assembled in our institute~\cite{Shin:2020bdq}.
Two custom-made Kyropoulos machines were implemented at the same site  for both the demonstration of low-background crystal growth
and the timely production of 200~kg of NaI(Tl) crystals~\cite{Ra:2018kkl}.
In the meantime, we developed detector assembly techniques for low-background NaI(Tl) detectors.
In this article, our progress in the development of low-background NaI(Tl) detectors and the prospects of the COSINE-200 experiment is described. 

\section{Crystal Growth}
\label{sec:1}

\begin{figure}[htbp]	
	\centering
	\includegraphics[width = 0.9 \columnwidth] {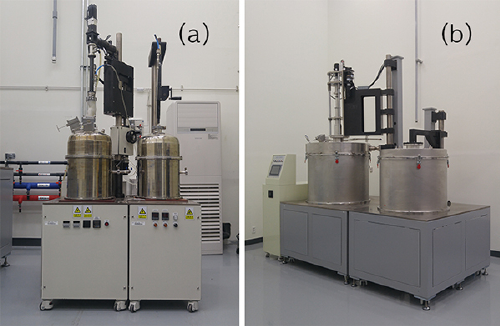}
	\caption{Two dedicated Kyropoulos growers for the NaI(Tl) crystals. (a) A small-sized grower for low-background NaI(Tl) R\&D and (b) a full-sized grower for 120~kg size NaI(Tl) crystal production. Both growers have an accompanying setup to
        facilitate hot crystal annealing.}
	\label{NaI_grower}
\end{figure}

\begin{figure}[htbp]	
  \centering
	\includegraphics[width = 0.5 \columnwidth] {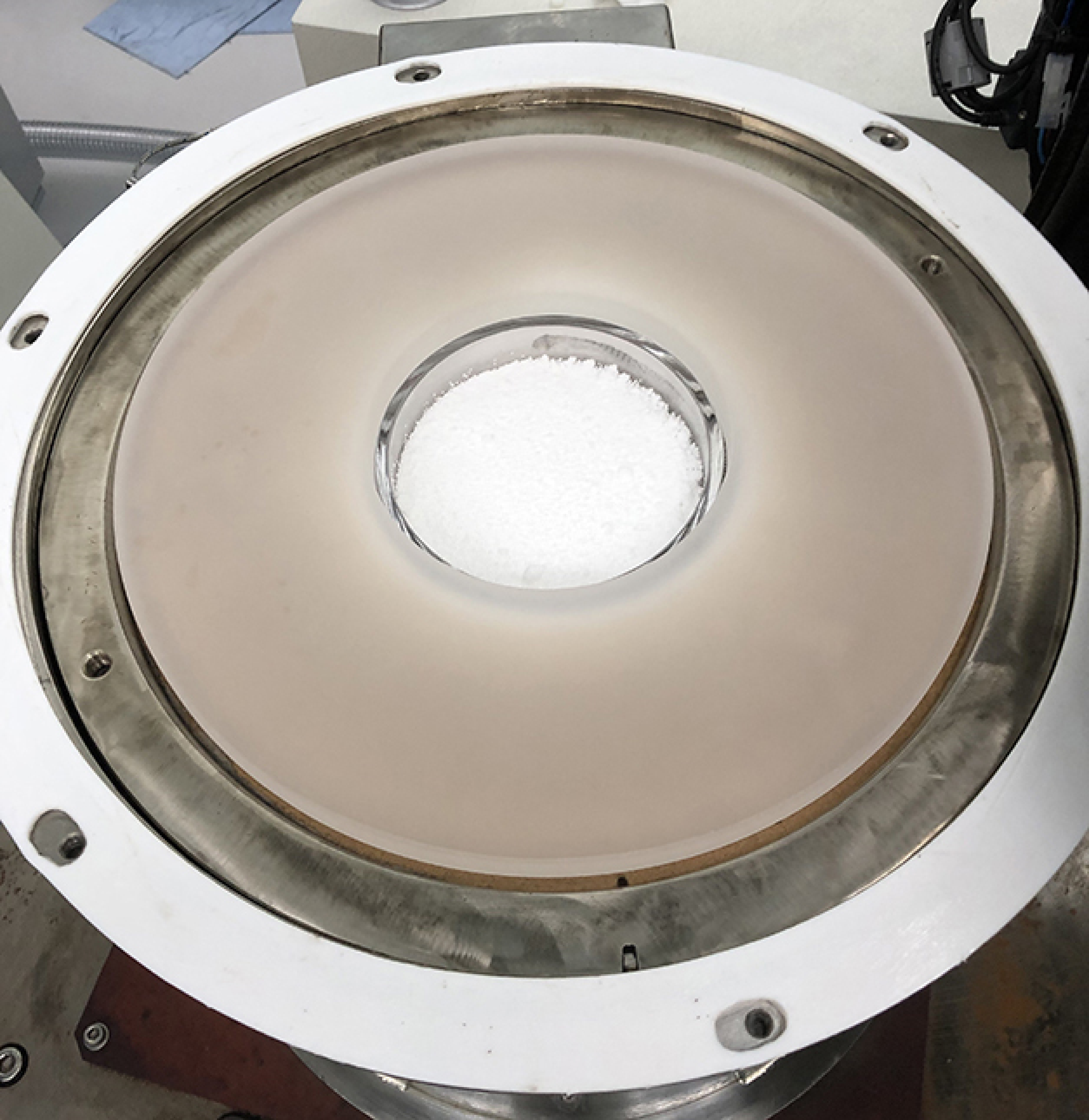}
        \caption{A top view of the grower with NaI powder loaded in the middle crucible,
          with the chamber cupola removed.
          A hollow-disk-shape quartz lid covers the top of the refractories to prevent fumes
          from coming out when heated. }
  \label{NaIquartz}
\end{figure}

We have built two dedicated Kyropoulos growers for NaI(Tl) crystals. A small R\&D grower that is used for proof of principle measurements is shown in Fig.~\ref{NaI_grower}(a). This grower is specifically designed to grow pilot NaI(Tl) crystals with relatively large temperature gradient and fast ingot rotation. 
The typical growth time is approximately 24~hours for a 70 mm diameter and 80 mm high ingot. 
Because of this short crystal growth time, we can  minimize radon contamination from emanations from the grower's materials and the ambient air. 

\begin{figure}[htbp]	
  \centering
  \includegraphics[width = 1.0 \columnwidth] {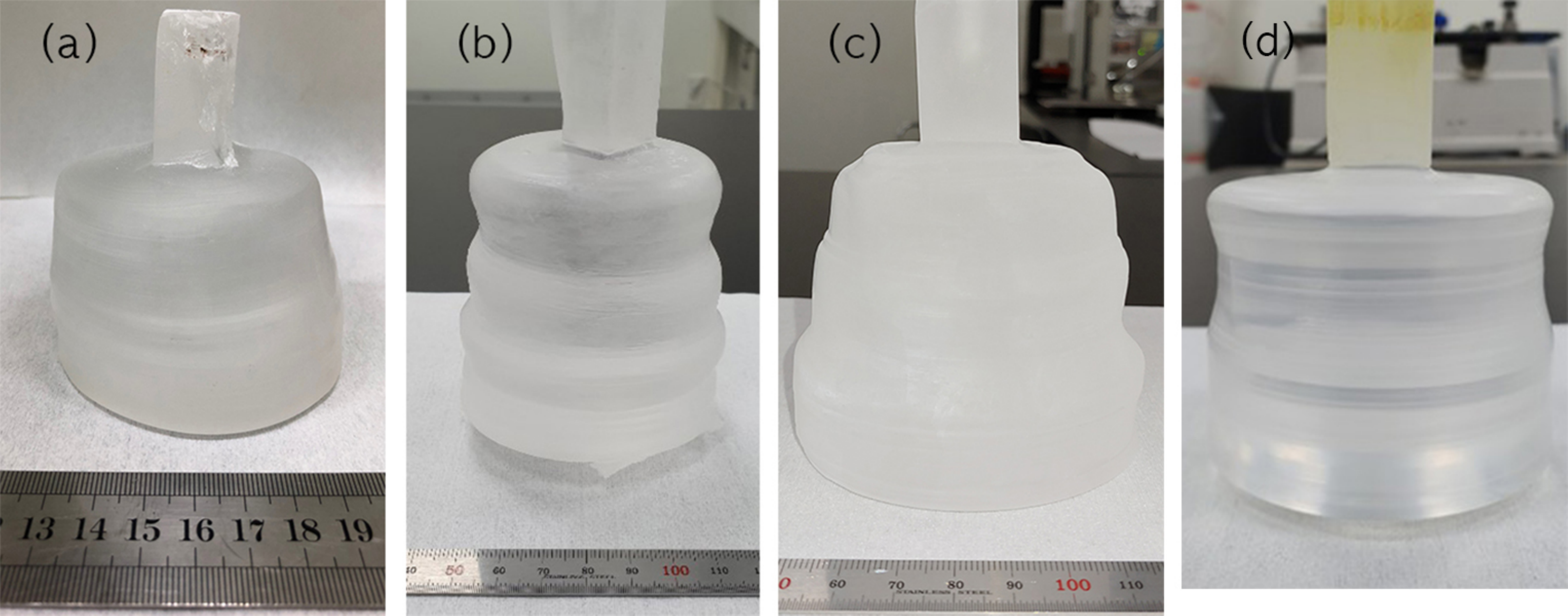}
  \caption{Four NaI(Tl) crystal ingots produced by the small grower : (a) NaI-025, (b) NaI-034, (c) NaI-035, and (d) NaI-036.}
  \label{NaIingot}
\end{figure}

For the small grower, we use a 12~cm diameter, 10~cm high quartz crucible that is surrounded by heaters and refractories inside the chamber. The chamber is cylindrical with dimensions of 34~cm diameter and 50~cm height. During crystal growth, N$_2$ gas was continuously flushed via a thallium trap with a flow rate of 10 liters/min. 
A same-size annealing machine is included as shown in Fig.~\ref{NaI_grower}(a). Figure~\ref{NaIquartz} shows the top view of the growing setup before the start the growth process. The quartz crucible can contain up to 1.7~kg of the NaI powder and can produce a 1.1~kg crystal ingot with typical dimensions of: diameter 60$-$80~mm height 50$-$80~mm as shown in Fig.~\ref{NaIingot}.   

With this system, we successfully produced several thallium-doped NaI(Tl) crystals with low-background NaI powder. Based on this success, we constructed a full-size grower that accommodates a 61~cm diameter and 44~cm height quartz crucible for the growth of approximately 120~kg crystal ingots (see Fig.~\ref{NaI_grower}(b)). The conceptual design of the full-size grower is similar to that of the smaller one, in order to minimize differences in growth conditions in terms of internal materials and thermal gradient. We expect approximately 3-4 days crystal growth times for 120~kg ingots (40 cm diameter and 25 cm height). 

All thallium-doped low-background crystals shown in Fig.~\ref{NaIingot} start with low-background NaI powder from the Sigma-Aldrich~(SA) company named ``Astro-Grade''. 
Previous studies~\cite{Adhikari:2017esn} show that this grade powder contains low potassium~(less than 10~ppb level) and lead~(a few ppb level).
Typically, the NaI powder from this vendor does not contain significant amounts of uranium and thorium. 

When the crystal growth is finished, sample pieces from the crystal ingot are measured with inductively coupled plasma mass spectrometry~(ICP-MS) to study impurity contaminations. The first grown crystal, NaI-025, was highly contaminated with both K and Pb, which was found to originate from fumes produced by the refractories as shown in Table~\ref{icpms}. 
The refractories were replaced with ones fabricated from relatively radiopure materials, and this resulted in significantly reduced impurities in the fumes. 
In addition, the implementation of a quartz plate that covers the top of the refractories prevented further diffusion of the fumes into the NaI crucible (see Fig.~\ref{NaIquartz}).
Subsequently grown crystals, NaI-034, NaI-035, and NaI-036, had
significantly reduced K and Pb contaminations, as can be seen in Table~\ref{crystal_growth}. 

\begin{table*} %[ht]
  \caption{ ICP-MS results for: initial materials; fumes produced during the growing process. 
    }
\label{icpms}
  \centering
\begin{tabular}{lcccc}
  \hline                                                                                
  Sample             & K (ppb) & Pb (ppb)  & U (ppt)  & Th (ppt)  \\
  \hline                 
  NaI (Astro-grade)  &  5       & 1.3      & $<$10    &  $<$10    \\
  TlI (SA powder)    &  3400    & 500      & $<$10000 & $<$10000  \\
  TlI (SA bead)      &   150    & 100      & $<$10000 & $<$10000    \\
  TlI (AA powder)    &  2100    & 600      & $<$10000    & $<$10000   \\
  TlI (AA bead)      &  4400    & 600      & $<$20000    & $<$10000   \\
  \hline                 
  Fumes with old refractories  & 1021235  & 10407     & $<$10   & $<$10 \\
  Fumes with new refractories  &  320 & 25     & $<$10   & $<$10     \\
  \hline  
\end{tabular}
\end{table*}

We tested different types and amounts of thallium iodide (TlI) from different vendors
as the dopant in the NaI(Tl) crystals.
Initially, we used 99.999\% purity TlI powder from SA.
However, this powder created black dust in the melt that was traced to carbon impurities in the powder.
This dust made it difficult to optimize crystal growth conditions and
the produced ingots contained black speckles throughout their entire bulk.  
We speculate that the carbon contamination in the TlI powder may have been caused
by plastic or polytetrafluoroethylene (PTFE) material used in the chemical purification process.  
After a variety of tests for different types of TlI, we found that the bead type TlI
does not produce any significant black dust either in the melt or the ingot. 
To confirm the reproducibility of this, we also tested TlI from Alpha-Aesar~(AA) in both powder and bead forms
and concluded that the bead type contains less carbon. 
Contamination levels of radioisotopes for various TlI samples  were measured with ICP-MS (see Table~\ref{icpms}).

Additionally, since the NaI(Tl) melt directly touches the crucibles,
we checked the impact of the crucible type by exchanging the quartz crucible with a platinum one. 
The platinum crucible provided a lower contamination of K and Pb although the crystals produced with quartz crucibles also had   $^{40}$K and $^{210}$Pb levels that are sufficiently low for the COSINE-200 experiment. 
Specifications of the crystal growth conditions and ICP-MS measurement results for various NaI(Tl) crystals grown in the laboratory
are summarized in Table~\ref{crystal_growth}. 

\begin{table*} %[ht]
  \caption{The top six entries list growth conditions of each crystal ingot, where SA and AA indicate Sigma-Aldrich and Alpha-Aesar, respectively. The middle two entries list  the  masses of the crystal ingots and polished detectors. The lower five entries list the ICP-MS results of the grown crystals, where top and bottom measurements are averaged and uncertainties are half the differences of the two measurements. }
\label{crystal_growth}
  \centering
\begin{tabular}{c|cccc}
  \hline                                                                                
  Crystal             & NaI-025 & NaI-034 & NaI-035 & NaI-036 \\
	\hline
  Growth date         & May/24/2018 & Sep/18/2019 & Nov/12/2019 & Feb/4/2020  \\
  Quartz cover        & no & yes & yes & yes \\ 
  Refractory          & old & new & new & new \\
  Crucible            & Quartz & Platinum & Quartz & Quartz \\  
	TlI loaded (mol \%) & 0.1 & 0.1 & 0.12 & 1.0 \\
	TlI powder          & SA powder & SA powder &AA bead & SA bead \\ 
  \hline                 
  Mass (ingot) (kg)   & 0.79 & 1.18 & 1.16 & 1.08 \\
	Mass (detector) (kg)& 0.65 & 0.67 & 0.61 & 0.78 \\
  \hline                 
	K (ppb)             & 740$\pm$68 & 8$\pm$3 & 13$\pm$6 & 23$\pm$9 \\
  Pb (ppb)            & 170$\pm$15 & 0.5$\pm$0.1 & 2$\pm$1 & 10$\pm$1 \\
	U (ppt)             & $<$10 &$<$10 &$<$10 &$<$10 \\
	Th (ppt)            & $<$10 &$<$10 &$<$10 &$<$10 \\
%	Tl (mol \%)         & 0.024$\pm$0.04 & 0.039$\pm$0.07 & 0.036$\pm$0.07 &0.197$\pm$0.014 \\
%	Tl (mol \%)         & 0.017$\pm$0.03 & 0.028$\pm$0.05 & 0.026$\pm$0.05 &0.141$\pm$0.010 \\
	Tl (mol \%)         & 0.008$\pm$0.001 & 0.015$\pm$0.001 & 0.023$\pm$0.001 &0.072$\pm$0.003 \\
  \hline  
\end{tabular}
\end{table*}

\section{Experimental setup}
\label{sec:2}

\subsection{NaI(Tl) crystal detectors} 

\begin{figure}[htbp]	
  \centering
  \includegraphics[width = 1.0 \columnwidth] {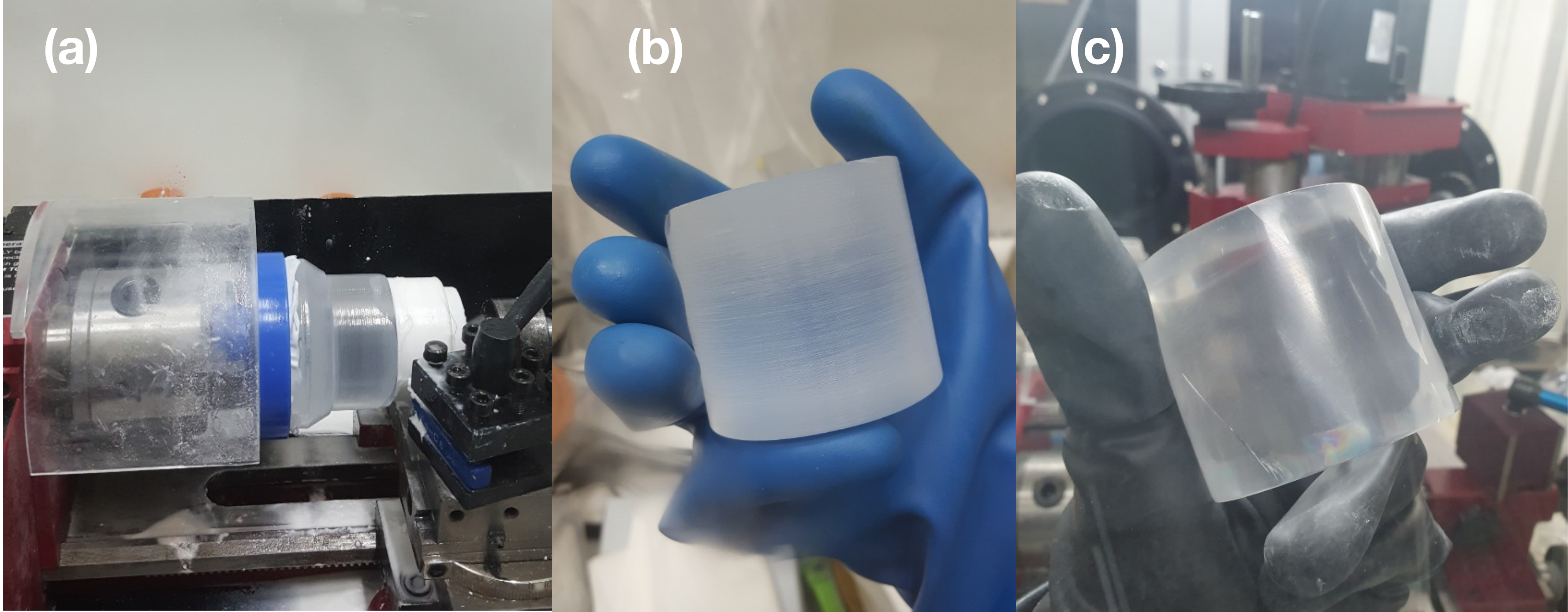}
  \caption{NaI-035 crystal machining and polishing.
    (a) The crystal is machined into a cylindrical form using a lathe.
  The finished product before (b) and after (c) the surface polishing is shown.}
  \label{NaIMach}
\end{figure}

Once the crystal ingots are prepared, samples from the top and bottom sections were cut out with a bandsaw for ICP-MS measurements. The remaining bulk of the crystal was machined to a cylindrical shape using a lathe in a low-humidity, N$_2$ gas flushed glovebox shown in Fig.~\ref{NaIMach}. While shaping the crystal, mineral oil was continuously poured on it to prevent cracks from developing and suppress NaI(Tl) dust in the environment. 
Four separate ingots were assembled as detectors after the crystal surfaces were polished, also in the low-humidity,  N$_2$ gas flushed glovebox. For NaI-025 and NaI-036, we did not machine the barrel surfaces, while NaI-034 and NaI-035 were machined into cylindrical forms. Studies show that the machined cylindrical surface improves light collection while not affecting the background contamination level.

\begin{figure*}[htbp]	
  \centering
  \includegraphics[width = 1.0 \textwidth] {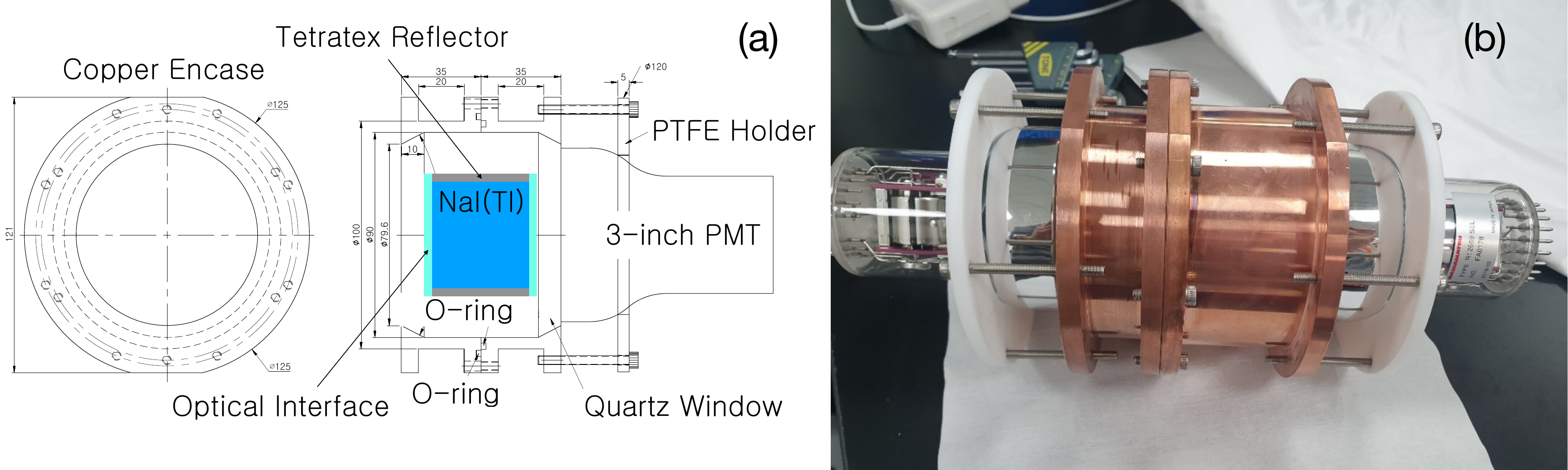}
  \caption{NaI-035 detector.
    The diagram (a) shows the encapsulation design and the photo (b) shows the final detector.}
  \label{Encap}
\end{figure*}

The crystal was dry-polished using aluminum oxide lapping films ranging
from 400 to 8,000 grits.
After the polishing, several layers of soft PTFE sheets
were wrapped on the barrel surface as a diffusive reflector
and an optical interface (silicon rubber EJ-560) was coupled to the top and bottom surfaces.
A 5 mm-thick cylindrical copper casing with quartz windows at each end
is used to hermetically encapsulate the crystal as shown in Fig.~\ref{Encap}.
Hamamatsu 3-inch PMTs (R12669SEL) are coupled via an optical interface to each quartz window.
The PMTs are fixed with PTFE guard rings by means of stainless-steel bolts.

All assembly procedures were performed inside the glovebox in which
the humidity level was maintained to be less than a few tens of ppm (H$_2$O) using
$\rm N_2$ gas and a molecular sieve trap.
All assembly parts were cleaned using diluted Citranox liquid with sonication,
and baked in an oven before they were moved  into the glovebox.
Several weeks after the assembly was finished, a visual inspection showed no degradation of the crystal
in terms of its transparency. For evaluation, we used an underground low-background setup for further detailed measurements.

\subsection{Shielding structure}
\begin{figure}[htbp]	
\centering
\includegraphics[width = 1.0 \columnwidth] {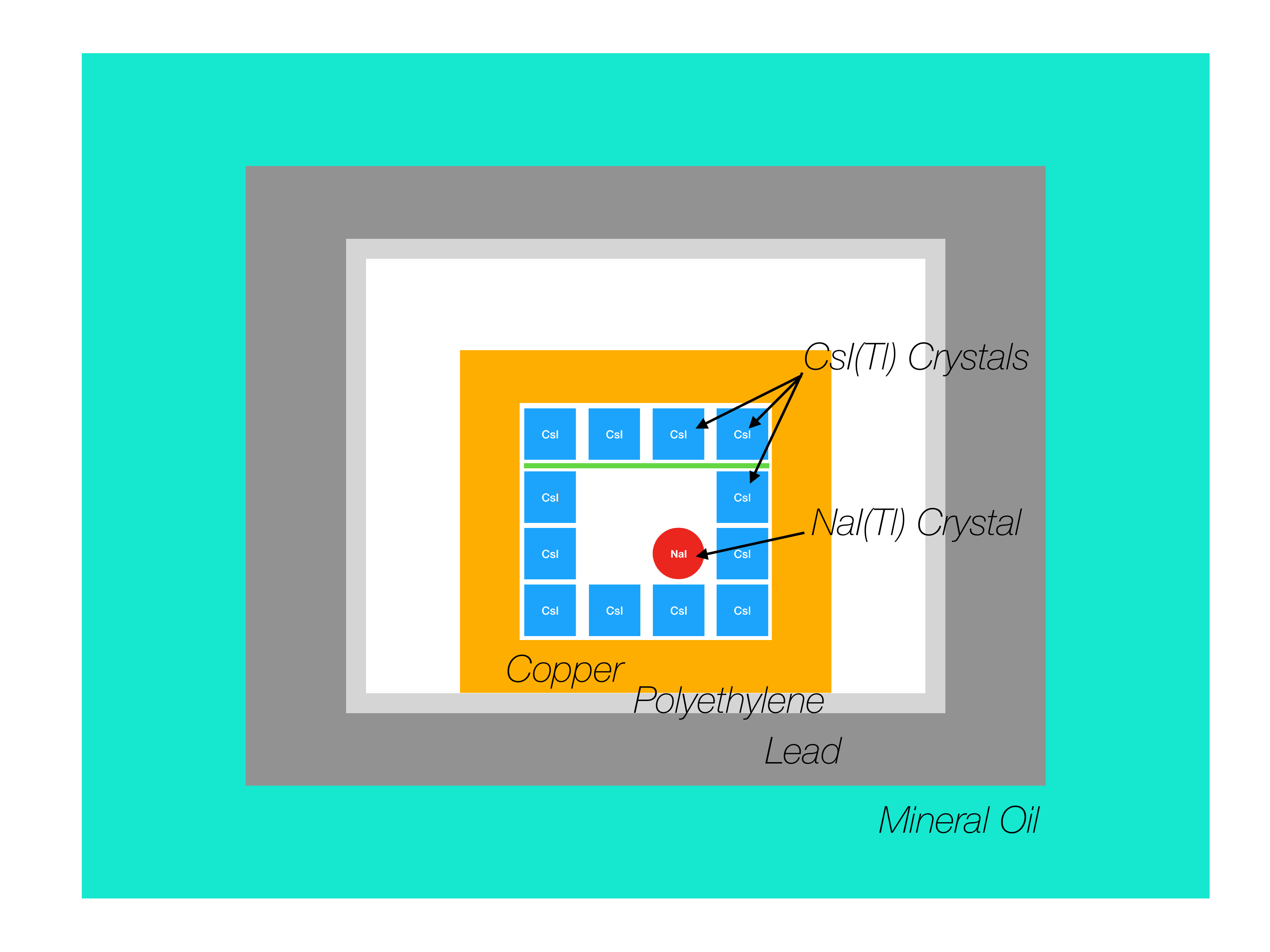}
\caption{A schematic view of the Y2L R\&D setup.
  The NaI(Tl) crystal (circle) is placed within the CsI(Tl) crystal array (squares)
  to tag external and coincident components of the testing crystal. An acrylic table (green solid line) supports top-side CsI(Tl) crystals. 
	These were installed inside a shield. 
  Each NaI(Tl) crystal was tested one by one in this setup.}
\label{a6setup}
\end{figure}

To evaluate the background contamination levels in the NaI(Tl) crystals, we used the same experimental apparatus
that was used for previous NaI(Tl) crystal R\&D at the Yangyang Underground Laboratory~(Y2L)~\cite{Kim:2014toa,adhikari16}.
This includes a 12-module array of CsI(Tl) crystals inside a
shield that is comprised of 10 cm of copper, 5 cm of polyethylene,
15 cm of lead, and 30 cm of liquid-scintillator-loaded mineral oil~\cite{Lee:2005qr,Lee:2007iq} as shown in Fig.~\ref{a6setup}. The shielding stops external neutrons and gamma
rays, and vetoes cosmic-ray muons.
The NaI(Tl) crystals were tested inside the CsI(Tl) detector array with an arrangement indicated schematically in Fig.~\ref{a6setup}. 

\subsection{Electronics}
Signals from PMTs attached at each end of the NaI(Tl) crystal are amplified by custom made preamplifiers: the high-gain anode signal by a factor of 30 and the low-gain signal from the 5th stage dynode by a factor of 100. The amplified signals are digitized by a 500~MHz, 12-bit flash analog-to-digital converters~(FADC). The FADC module has four channels that accommodates one NaI(Tl) crystal's data. Triggers from individual channels are generated by field programmable gate arrays~(FPGAs) embedded in the FADC. The final trigger decision for an event is made by a trigger and clock board (TCB) that also synchronizes the timing of the different modules. 
A trigger is generated when an anode signal corresponding to one or more photoelectrons (PEs) occur in each PMT within a 200 ns time window. 
Both the low-gain dynode and high-gain anode signals are recorded whenever the anode signals produce a trigger. The dynode signals do not produce triggers. The dynamic range of the anode signals is up to 100~keV while that for the dynode signals is up to 5000~keV. 

Signals from the CsI(Tl) crystals are amplified by a gain of 10 and digitized in a charge-sensitive 62.5 MHz ADC (SADC). The SADC returns the integrated charge and the time of the signal. We use a 2048~ns integration time to contain CsI(Tl) signals considering their decay times. The SADC channels do not generate triggers. 

If the NaI(Tl) crystal satisfies the trigger condition, the TCB transmits trigger signals to the FADC and the SADC. Data of the NaI(Tl) crystal and the CsI(Tl) crystals are stored for the FADC trigger. For the FADC channel, this corresponds to an 8~$\mu$s-long waveform staring approximately 2.4~$\mu$s before the time of the trigger position. For each SADC channel this corresponds to the maximum integrated charge within an 8~$\mu$s search window and to the associated time of that maximum. 
This system has been used in COSINE-100 experiment and descriptions are provided elsewhere~\cite{Adhikari:2017esn,Adhikari:2018fpo}.  

\subsection{Energy calibration and light yields}
The low energy anode signals of the NaI(Tl) crystals are calibrated with 59.54~keV $\gamma$-rays emitted from an $^{241}$Am source. 
The high energy dynode signals are calibrated with 609~keV ($^{214}$Bi) and 1460~keV ($^{40}$K) lines that are produced by contaminants in the crystal. 

The charge distribution of single PE signals is obtained by identifying isolated clusters in the decay tail of the 59.54~keV $\gamma$-ray signal (2$-$4 $\mu$s after the signal start). 
From this, the light yield is determined from the ratio of the total deposited charge to the single PE's mean charge, scaled to 59.54~keV. 

A high NaI(Tl) crystal light yield is crucial for enabling the COSINE-200 experiment
to operate with a 1 keV energy threshold~\cite{Adhikari:2020xxj} that matches the DAMA experiment's conditions~\cite{Bernabei:2018yyw}. 
The doping concentration of thallium
is one of the most important factors that determine the light yield, as described in Refs.~\cite{doi:10.1143/JPSJ.69.3435,TREFILOVA2002474}.
We originally targeted a doping level of 0.1~mol\% for an optimal light yield based on the above-noted references. 
To accomplish this, we applied approximately 0.1~mol\% TlI powder into the initial mix for the NaI-025, NaI-034, and NaI-035 crystals. 
However, the measured doping levels in the final produced crystals were significantly reduced to 0.01$-$0.02~mol\% (Table~\ref{icpms}). 
This was mainly attributable to the evaporation of a large fraction of the thallium during the high-temperature growth period. 
We measured relatively low light yields from these crystals; between 9 and 12 number of PEs (NPEs) per keV. 
Because of light absorption by the black impurities inside the crystals, slightly lower light yields in the crystals grown with  TlI powder (NaI-025 and NaI-034) were observed (see Table~\ref{table:crystals}). 

For NaI-036, 
1~mol\% of bead-type TlI from SA was added to the initial mix and a 0.07~mol\% Tl concentration was measured in the ingot after growth.
In this crystal, we obtain the highest light yield of 17.1$\pm$0.5 NPE/keV.
Further increases in the light yield are possible with the improved encapsulation scheme that is described in Ref.~\cite{Choi:2020qcj}. 
Note that the COSINE-100 crystals and DAMA crystals have light yields of approximately 15 NPE/keV and 5$-$10 NPE/keV, respectively, 
and are able to operate with a 1~keV threshold~\cite{Bernabei:2018yyw,Adhikari:2020xxj}.
If the optical quality in crystals from the full-sized grower can match that of NaI-036, the light yield will be sufficient for operation of the COSINE-200 experiment
with a 1~keV energy threshold. 

\subsection{PMT noise background and low-energy event selection}
Photomultiplier tubes generate low-energy noise signals that are produced by a variety of mechanisms~\cite{Kim:2014toa,Bernabei:2008yh}. 
The DAMA group reported a signal selection criterion for efficiently removing 
PMT noise that exploits the  fast decay time characteristics of PMT noise pulses. The ratio of ``fast'' charge (0-50~ns) and ``slow'' charge (100-600~ns) is used as a single parameter to discriminate
fast PMT-induced noise pulse from genuine scintillation-induced signals~\cite{Bernabei:2008yh,Bernabei:2012zzb}. In addition, our COSINE-100 experience showed that further selections are required based on 
the asymmetry between the two PMT signals and the average charge of the isolated clusters that comprise each signal~\cite{Kim:2014toa,Adhikari:2017esn}. 
A further optimized selection using a multivariable machine-learning algorithm, the boosted decision trees~\cite{BDT}, was developed for the COSINE-100 experiment~\cite{Adhikari:2018ljm,Adhikari:2020xxj}. 
However, we only applied three parameter-based selection criteria discussed in Refs.~\cite{Kim:2014toa,Adhikari:2017esn} because the short-term measurements of the R\&D data were not sufficient to produce adequate training samples. 

\section{Background understanding} 
\label{background}
\subsection{Internal natural backgrounds}
To produce ultra-low-background NaI(Tl) crystals, one should understand the internal contamination of natural radioisotopes. 
Contamination levels for various radioisotopes are described for each of the four crystals below.
Table~\ref{table:crystals} shows a summary of the internal background measurements that are discussed in this section. 
The results are compared with the DAMA~\cite{Bernabei:2008yh,Bernabei:2012zzb}, and the SABRE~\cite{Suerfu:2019snq} crystals, and crystal-6 (C6) of the COSINE-100 experiment~\cite{cosinebg}. (C6 has the lowest background level among all the COSINE-100 crystals.)

\begin{table*} %[ht]
		\caption{Measured results of detector masses, light yield (LY), and the internal contaminations from this study compared with C6 of COSINE-100~\cite{cosinebg}, DAMA crystals~\cite{Bernabei:2008yh,Bernabei:2012zzb}, and SABRE result~\cite{Suerfu:2019snq}. }
\label{table:crystals}
\begin{tabular}{c|c|cccc|cccc}
  \hline                                                                                
	\multicolumn{2}{c|}{Crystal}         & NaI-025& NaI-034& NaI-035& NaI-036 &COSINE-100 & DAMA & SABRE \\
	\hline
	\multicolumn{2}{c|}{Mass (kg)} & 0.65 & 0.67 & 0.61 & 0.78 & 12.5 & 9.7 & 3.4 \\ 
	\multicolumn{2}{c|}{LY~(NPE/keV)}    & 10.4$\pm$1.6 & 9.5$\pm$1.1  & 11.8$\pm$1.8 & 17.1$\pm$0.5 & 14.6$\pm$1.5  & 5-10 & 16.4$\pm$0.3 \\
	\hline
	\multicolumn{2}{c|}{$^{nat}$K (ppb)} & 684$\pm$100 & $<$62  &   $<$42   &  $<$53 & 17$\pm$3 & $<$20 & 4.3$\pm$0.2 \\
  \hline  
	\multicolumn{2}{c|}{$^{210}$Pb (mBq/kg)} & 3.8$\pm$0.3   &0.05$\pm$0.09 & 0.01$\pm$0.02   & 0.42$\pm$0.27 & 1.87$\pm$0.09  &0.01$-$0.03&0.34$\pm$0.04  \\
  \hline  
	\multirow{3}{*}{ $^{238}$U} & $^{214}$Po ($\mu$Bq/kg) & 26$\pm$7   & 46$\pm$7   & 13$\pm$6   & 451$\pm$48  &$<$0.25 &$-$ &$-$  \\
   & $^{218}$Po ($\mu$Bq/kg)  & $-$       & 64$\pm$11 & 8$\pm$6   & $-$      &$-$ &$-$ &$-$  \\
	 & Average ($\mu$Bq/kg) & 26$\pm$7     & 51$\pm$7     & 11$\pm$4     & 451$\pm$48   & $<$0.25 & 8.7$-$124   &$-$  \\
  \hline  
  $^{232}$Th  & $^{216}$Po ($\mu$Bq/kg) & $<$6 & 35$\pm$5& 7$\pm$2 & $<$20   &2.5$\pm$0.8 &2$-$31 &$-$  \\
  \hline  
\end{tabular}
\end{table*}

\subsubsection{$^{40}$K background}
One of the most troublesome sources of background for WIMP searches with NaI(Tl) crystals is contamination from $^{40}$K.
Its natural abundance is roughly 0.012~\% of the total amount of potassium ($^{nat}$K). About 10~\% of $^{40}$K decays produce a
1460 keV $\gamma$-ray in coincidence with a 3.2~keV X-ray. If the
1460 keV $\gamma$-ray escapes the crystal and only the 3.2~keV X-ray is detected, an event is produced that is similar in energy
to that expected for a WIMP-nuclei interaction~\cite{cosinebg,Adhikari:2017gbj,Amare:2018ndh}. The $^{nat}$K contents in the DAMA crystals are in the 10$-$20 ppb
range~\cite{Bernabei:2008yh}, for the SABRE crystal it is 4.3$\pm$0.2 ppb \cite{Suerfu:2019snq}, and for the COSINE-100 C6 it is  16.8$\pm$2.5 ppb \cite{cosinebg}. 

In our apparatus~(see Fig.~\ref{a6setup}), $^{40}$K decays can be identified by coincidences between 1460 keV  $\gamma$-rays in the CsI(Tl) detectors and 3.2~keV X-rays in the tested NaI(Tl) crystal. 
The $^{40}$K background
level in each crystal is determined by comparing the measured coincidence rate with a GEANT4-simulated rate using
the method described in Refs.~\cite{Kim:2014toa,adhikari16}.

\begin{figure*}[htbp]	
	\centering
	\begin{tabular}{cccc}
	\includegraphics[width=0.24 \textwidth] {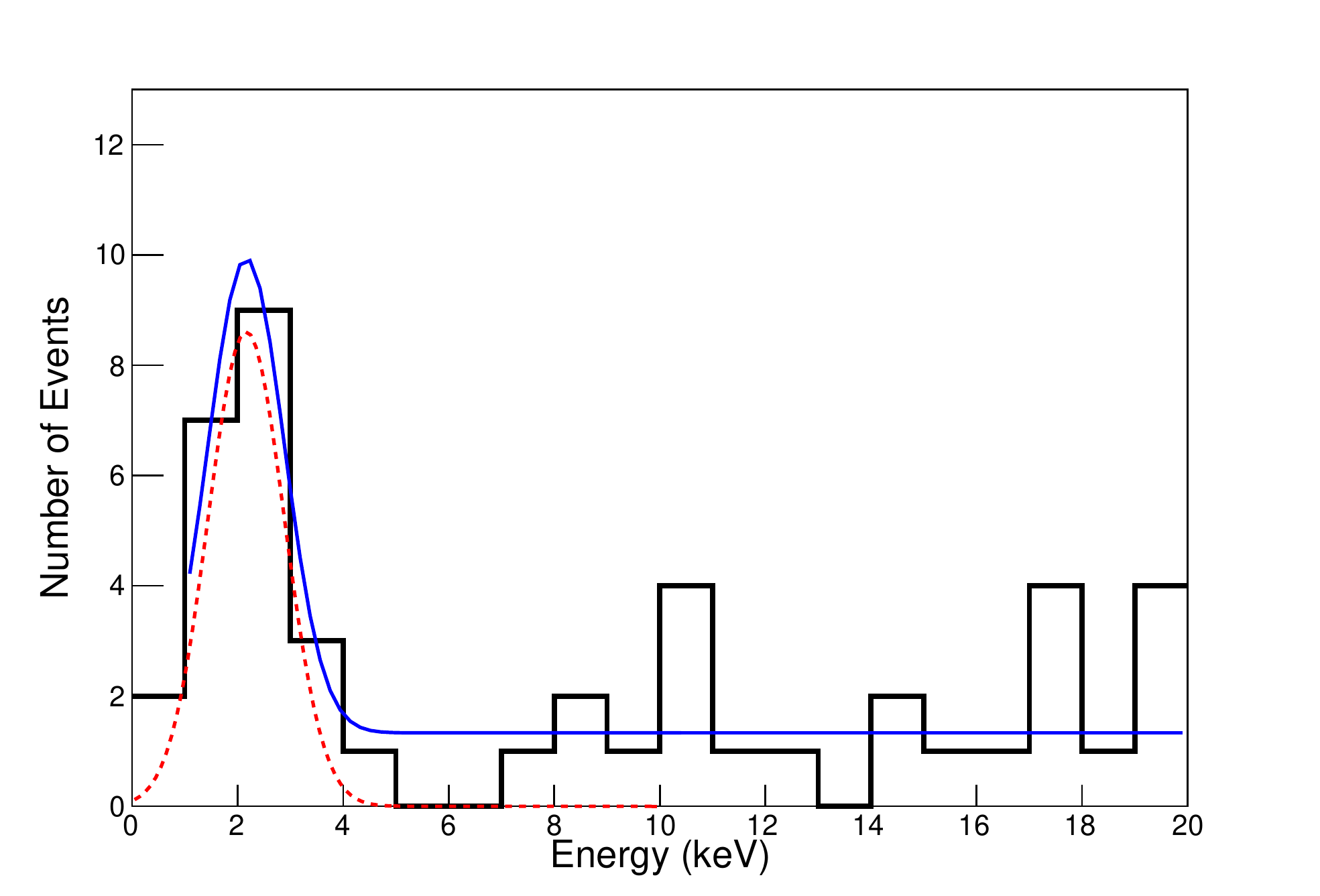} &
	\includegraphics[width=0.24 \textwidth] {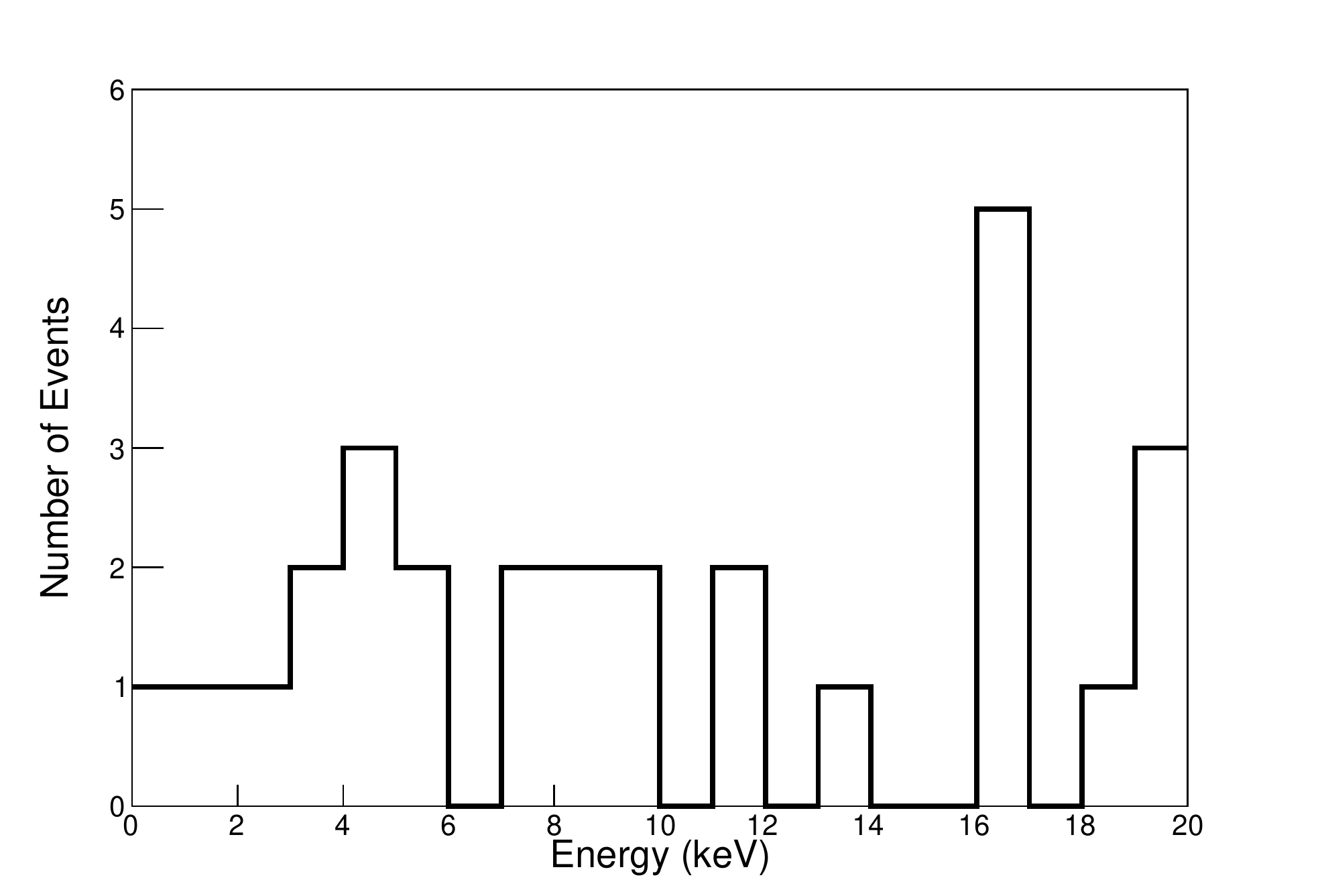} &
	\includegraphics[width=0.24 \textwidth] {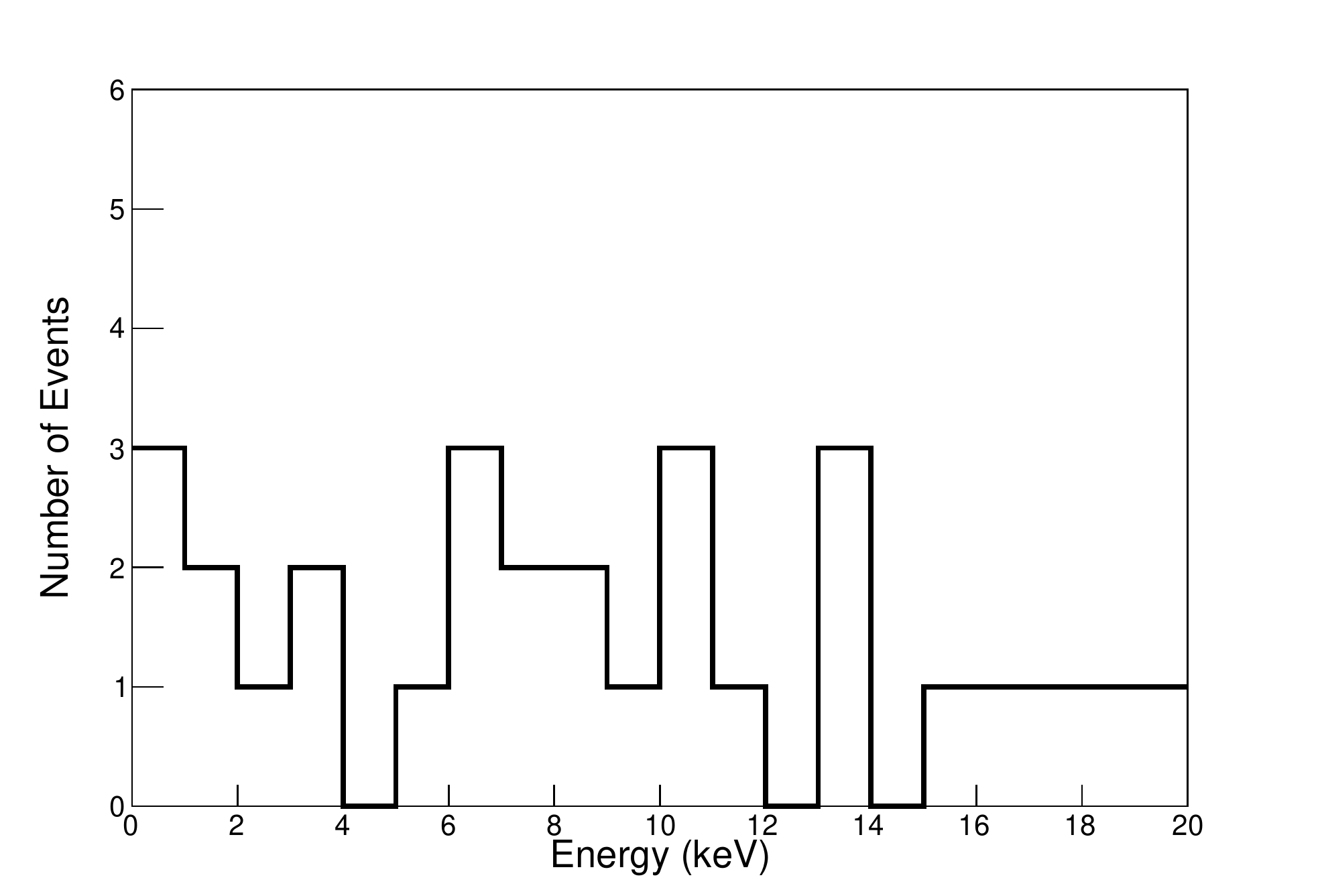} &	
	\includegraphics[width=0.24 \textwidth] {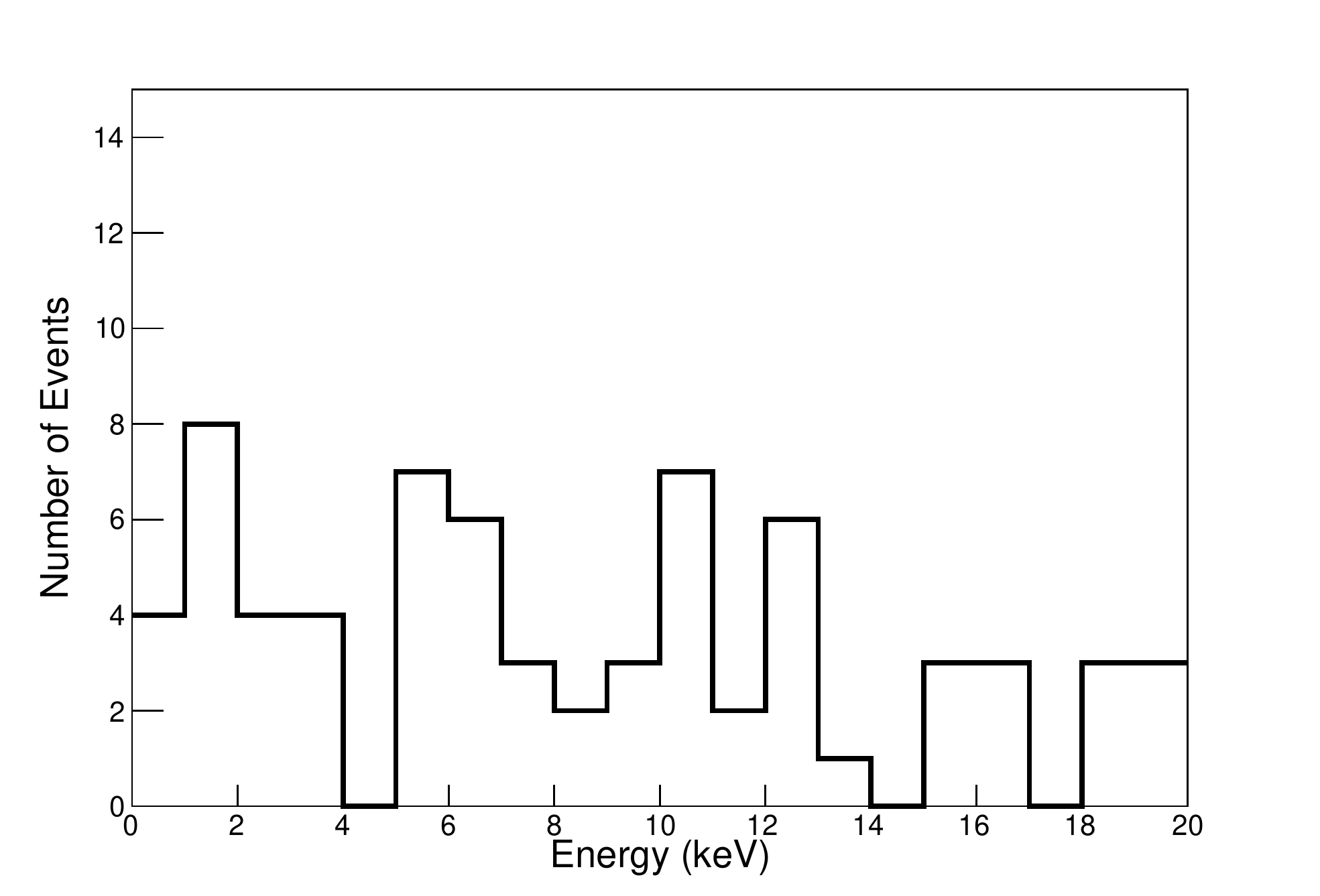}   \\
	(a) NaI-025 (16 days) & (b) NaI-034 (17 days) & (c) NaI-035 (23 days) & (d) NaI-036 (16 days)\\
	\end{tabular}
	\caption{The $^{40}$K 3.2~keV spectra in the NaI(Tl) crystals tagged by a 1460~keV $\gamma$ in the surrounding CsI(Tl) crystals. (a) A distinct 3.2~keV X-ray peak in NaI-025 reflects its high contamination of K. We model the energy spectrum~(solid line) assuming a Gaussian $^{40}$K signal~(red-dotted line) with a flat background.  However, there is no evidence for the 3.2~keV line in NaI-034 (b) and NaI-035 (c) and NaI036 (d). The exposure time for each crystal is denoted in parentheses. } 
	\label{potassium}
\end{figure*}

Figure~\ref{potassium} shows low energy spectra of the NaI(Tl) crystals associated with a 1460~keV $\gamma$ tagged in the surrounding CsI(Tl) crystals. A clear 3.2~keV peak in NaI-025 (Fig.~\ref{potassium} (a))  corresponds to a contamination of 684$\pm$100~ppb $^{nat}$K. 
In contrast, we found no evidence for potassium in NaI-035 and set a limit of 42~ppb~(90\% confidence level). 
Similarly, $<$62~ppb and $<$53~ppb are obtained for  NaI-034 and NaI-036, respectively, which are consistent with the ICP-MS measurements in Table~\ref{crystal_growth}.
Due to the small size of the tested crystals and the brief, one-month long measurement period, there are large uncertainties in the $^{nat}$K measurements.
Because the ICP-MS results provided consistent and much precise results on $^{nat}$K, we use the results from the ICP-MS measurements for the analysis discussed below. 

\subsubsection{$\alpha$ analysis}
\begin{figure}[htbp]	
  \centering
  \begin{tabular}{cc}
    \includegraphics[width=0.49 \columnwidth] {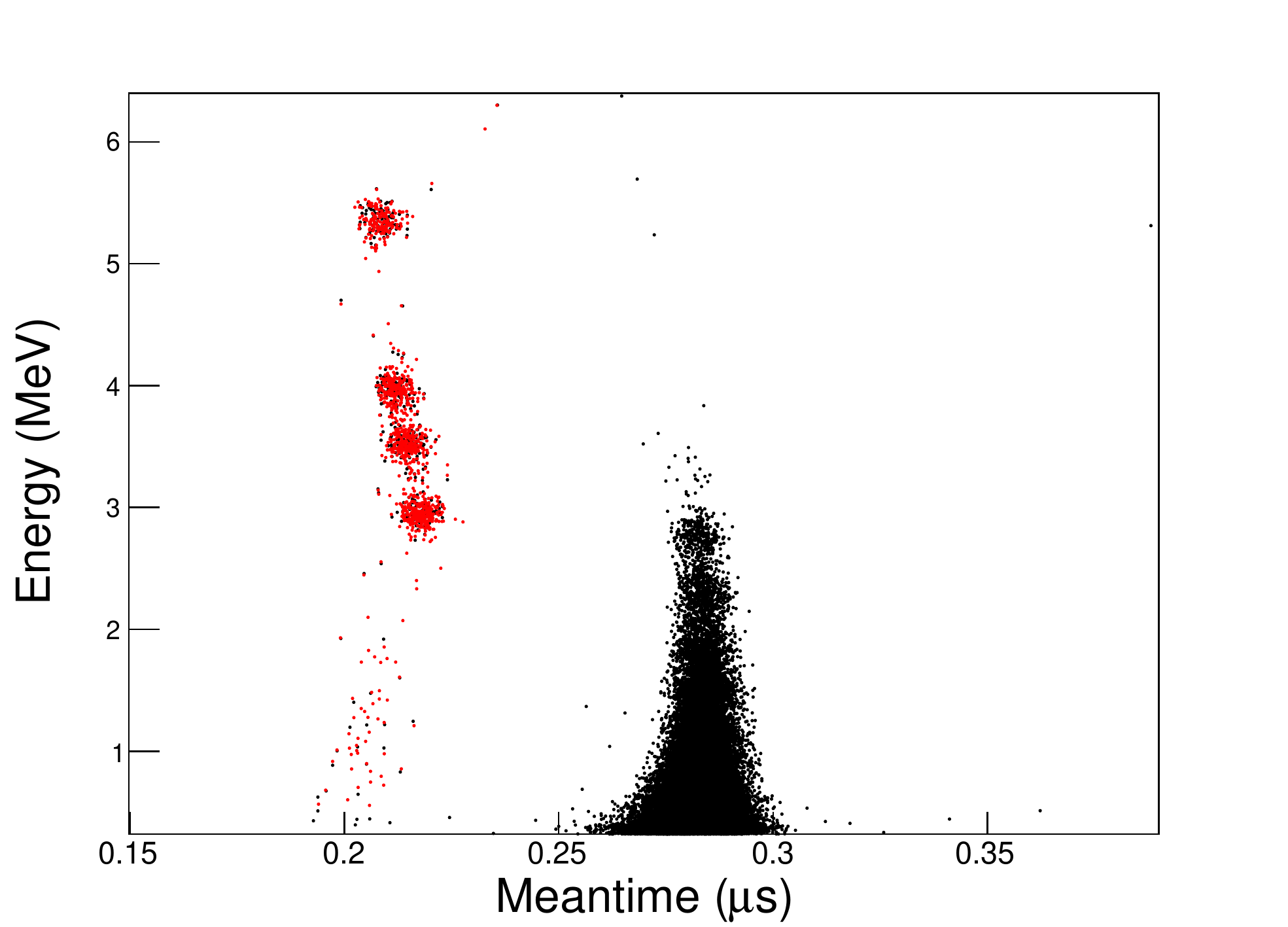}&
    \includegraphics[width=0.49 \columnwidth] {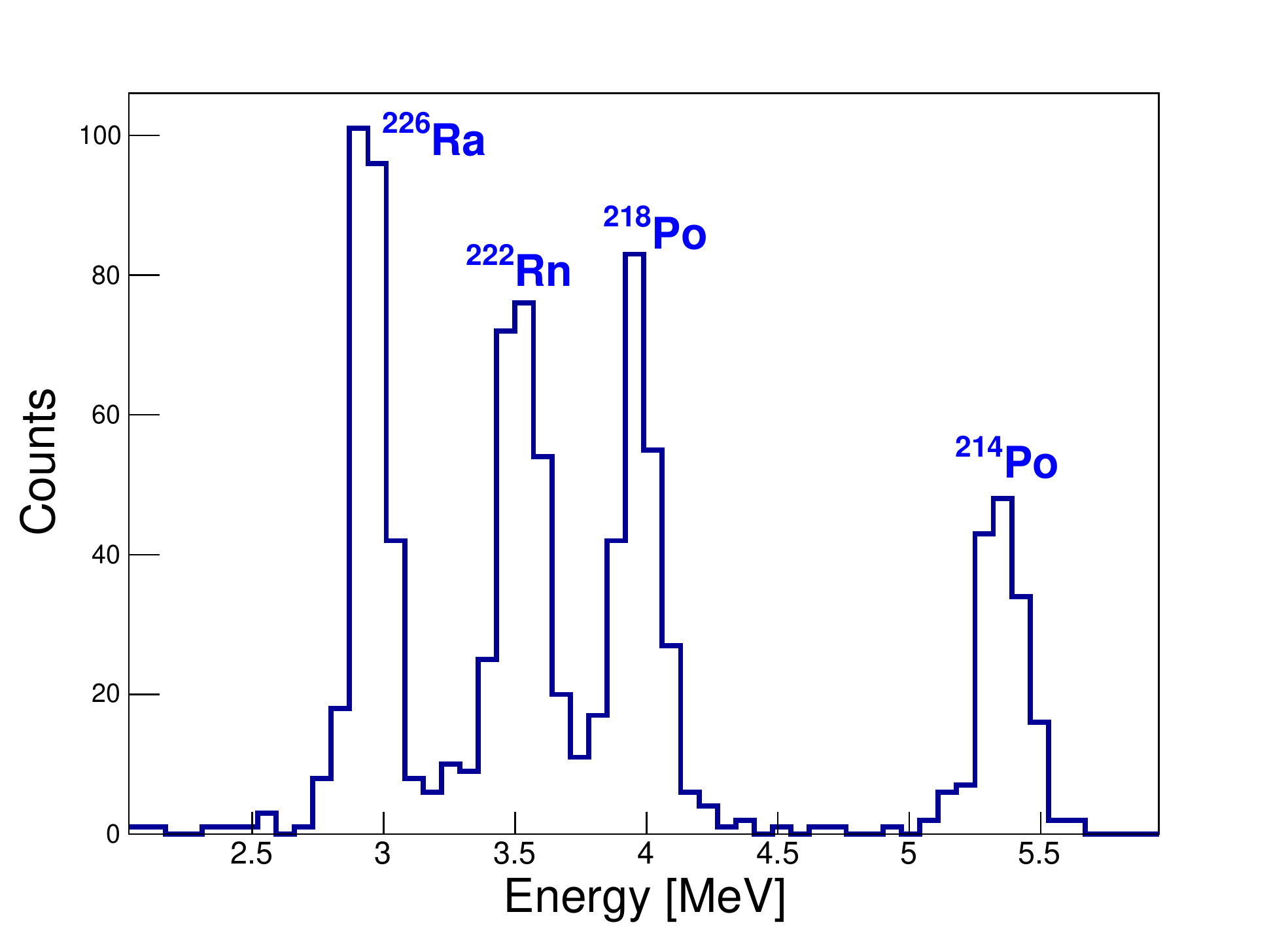} \\
    (a) & (b) \\
  \end{tabular}
  \caption{A scatter plot of the mean time versus the measured energy for the NaI-036 crystal with 16 days exposure is shown (a).
    Here $\alpha$-induced events (red dots) are well separated from $\beta$ or $\gamma$-induced events (black dots). 
    (b) The energy spectrum of the selected $\alpha$-induced events is shown; each peak represents
    a specific daughter alpha decaying component from $^{226}$Ra.
  }		
  \label{fig:alpha}
\end{figure}

Alpha-induced events inside the crystals can be identified by the mean time of their signals, defined as,
$$ <t> = \frac{\Sigma_i A_i t_i}{\Sigma_i A_i}.$$
Here $A_i$ and $t_i$ are the charge and time in each digitized bin of a recorded event waveform. Figure~\ref{fig:alpha}(a) shows a scatter plot of the energy versus the mean time for the NaI-036 crystal. Alpha-induced events are clearly separated from $\beta$ or $\gamma$-induced events because of their faster decay times. Selected $\alpha$ events (red points) are used to understand the internal contamination of heavy elements such as $^{238}$U, $^{232}$Th, and $^{210}$Pb. 
We also analyzed the energy spectra of these $\alpha$ events~(see Fig.~\ref{fig:alpha}(b)) in order to understand the contamination. Results of our analysis of the $\alpha$ contaminations are summarized in Table~\ref{table:crystals}.

\subsubsection{$^{210}$Pb background}

\begin{figure*}[htbp]	
	\centering
	\begin{tabular}{cccc}
	\includegraphics[width=0.24 \textwidth] {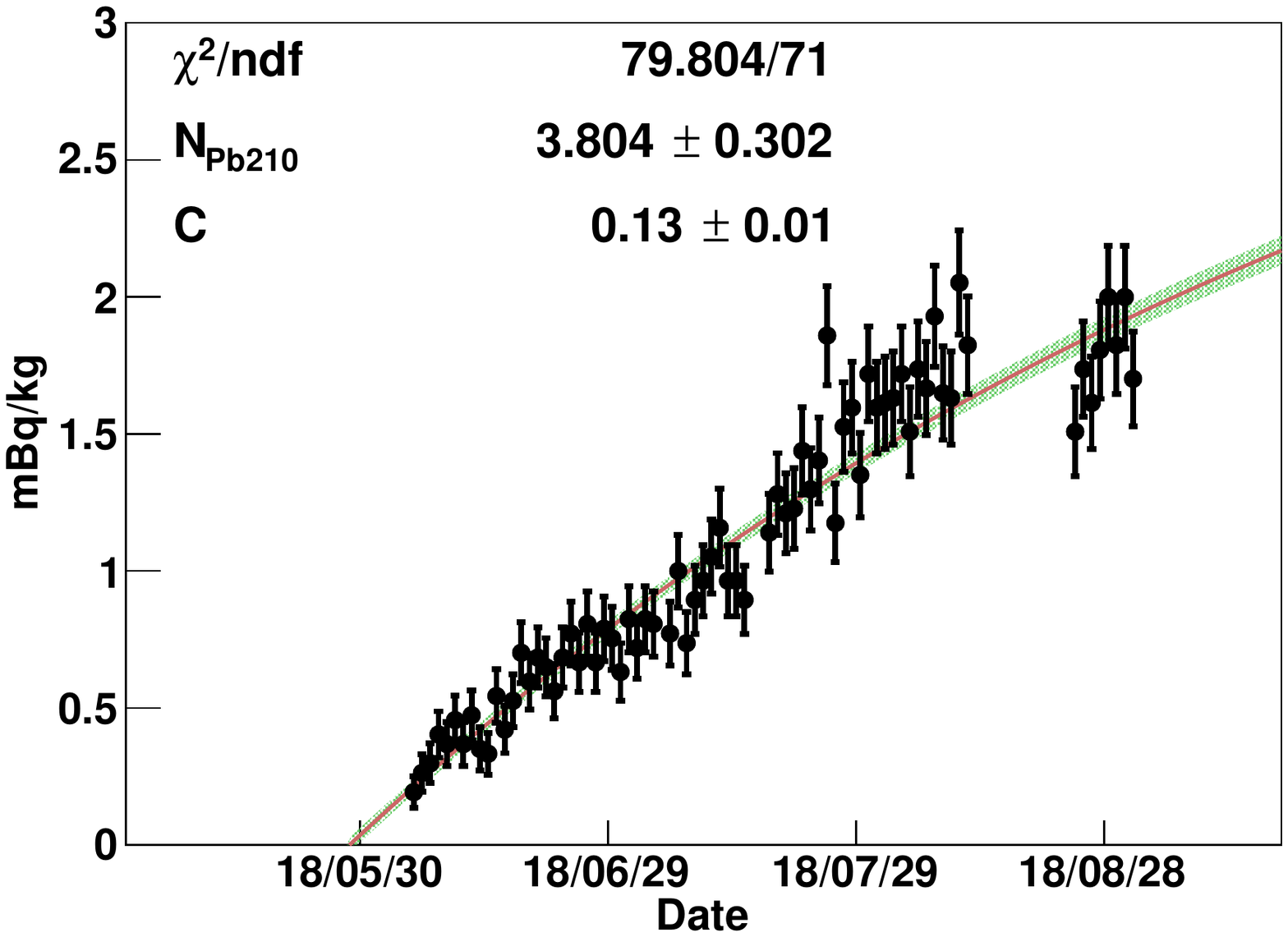}  &
	\includegraphics[width=0.24 \textwidth] {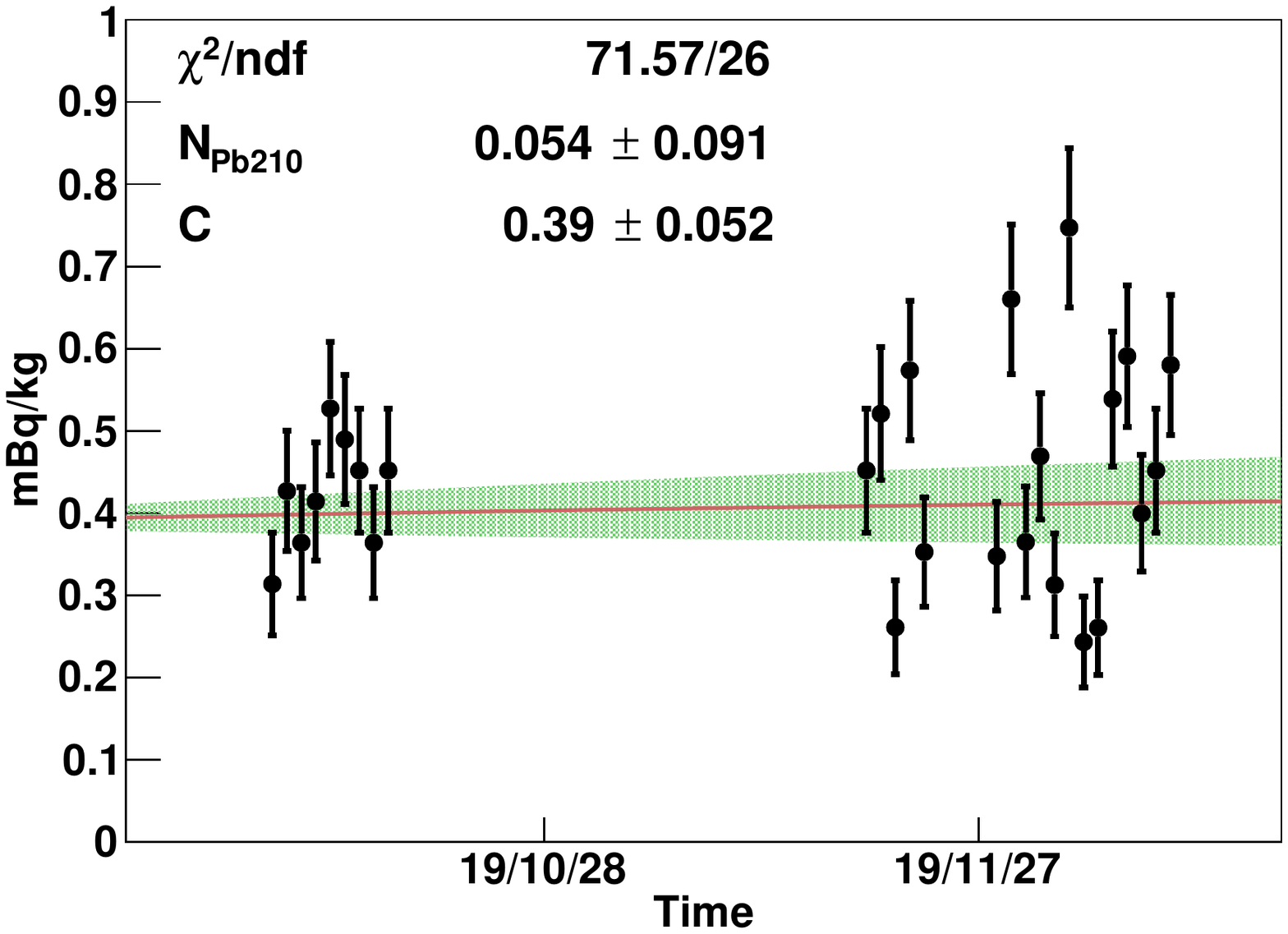} & 
	\includegraphics[width=0.24 \textwidth] {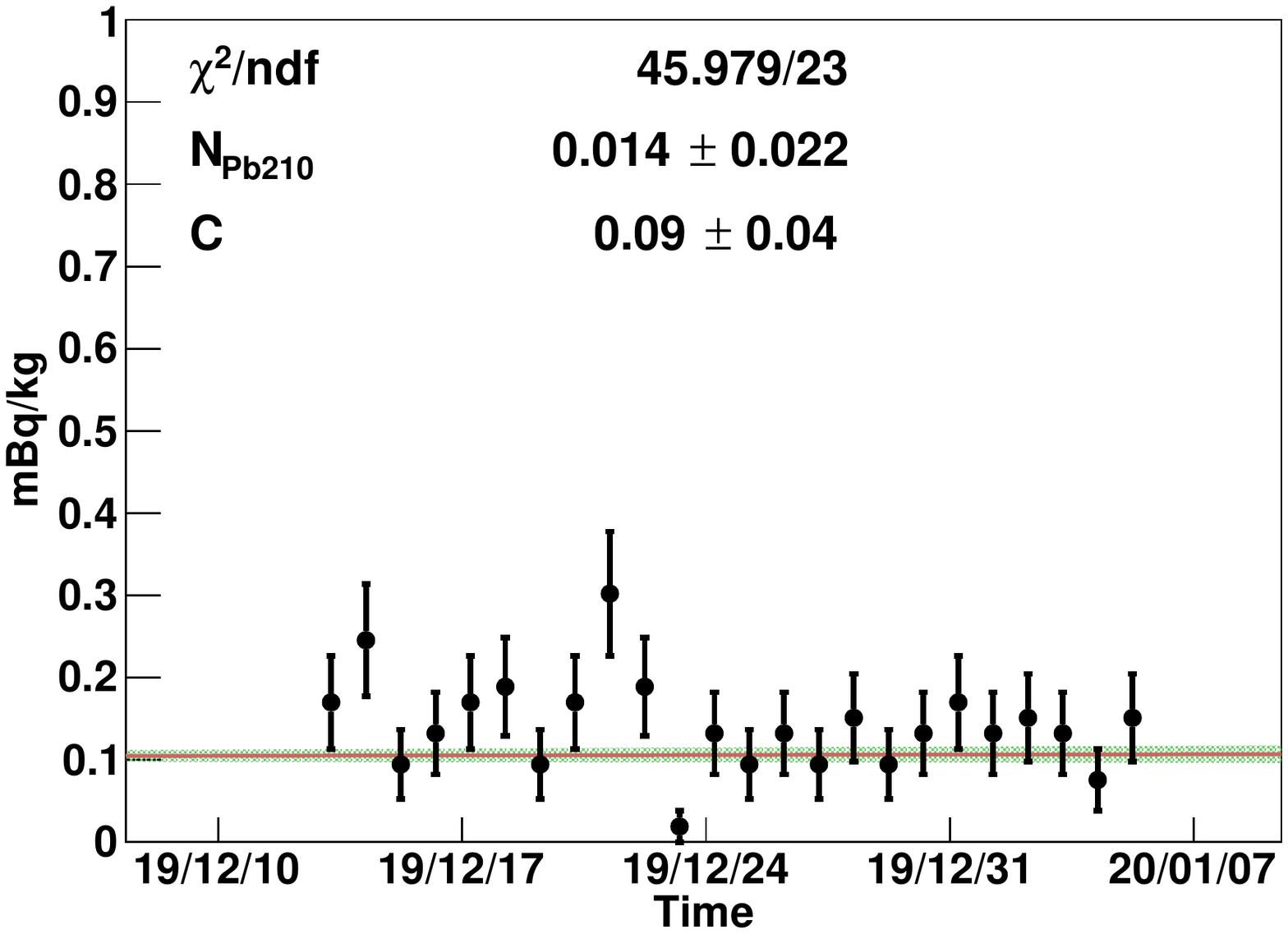} &
	\includegraphics[width=0.24 \textwidth] {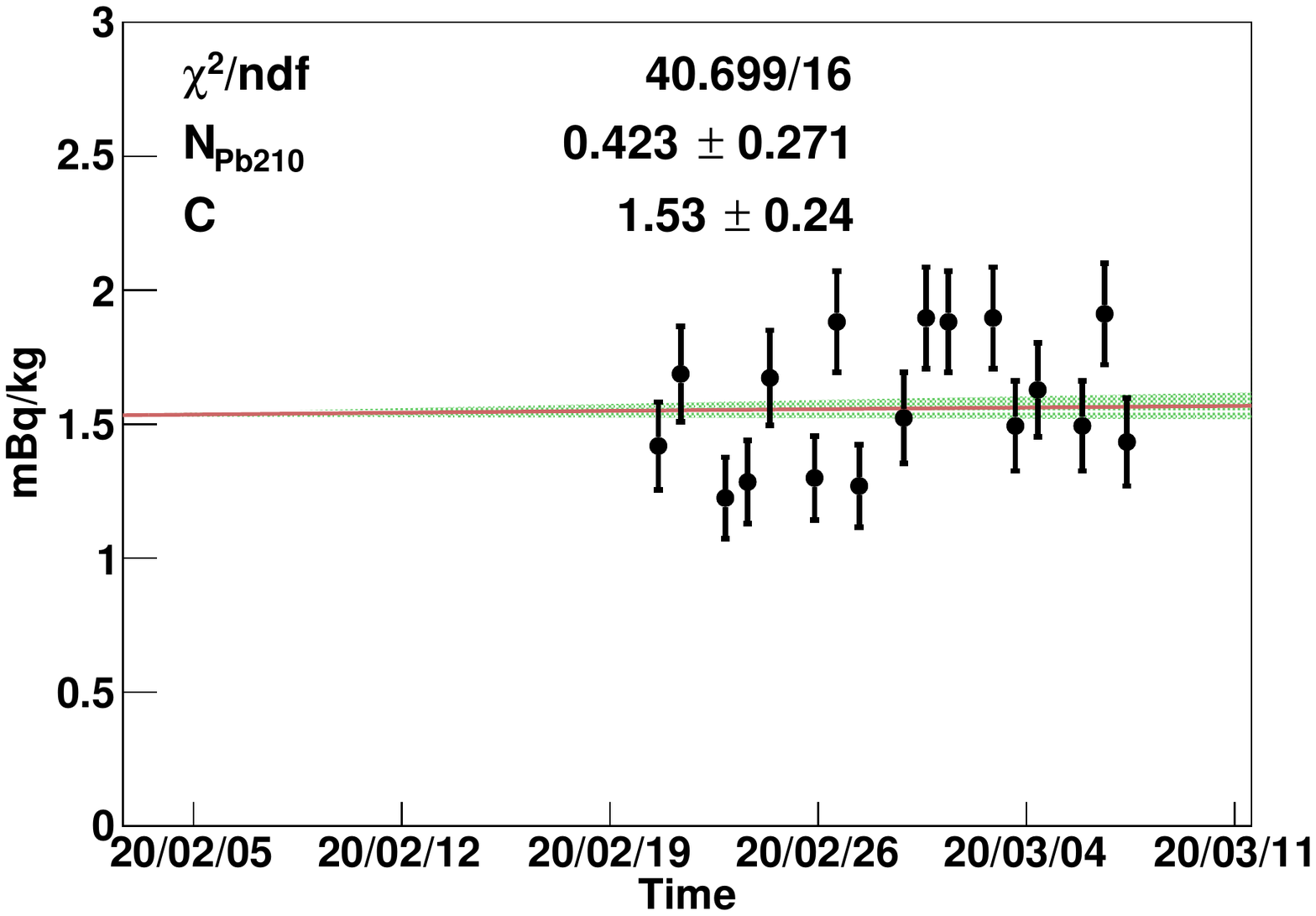} \\
	(a) NaI-025 & (b) NaI-034 & (c) NaI-035 & (d) NaI-036 \\
	\end{tabular}
	\caption{The total alpha rates in the crystals as a function of  time are modeled as a contamination of $^{222}$Rn (and/or $^{210}$Pb) at the crystal growth dates described in the text. Here black points are data, red-solid lines are the best fit models, and green bands are 1$\sigma$ uncertainty regions of the fit. %Half-life of $^{210}$Po is fixed as 138 days.
			(a) An increase in the alpha rates corresponding to the $^{210}$Po lifetime is evident in NaI-025.  
			No significant increases in alpha rates indicate that there are no significant $^{210}$Pb contaminations present in NaI-034 (b) or NaI-035 (c) during the crystal growth. 
			(d) $^{226}$Ra contamination in NaI-036 resulted in a relatively large number of measured alpha events. The total alpha rate is consistent with a 0.45~mBq/kg contamination of $^{210}$Pb that originated from $^{226}$Ra decays.}
	\label{alphatime}
\end{figure*}

In the COSINE-100 experiment, the main source of the alpha contamination was due to decays of $^{210}$Po (E$_{\alpha}$ = 5.4 MeV) nuclei that originated from a $^{222}$Rn contamination~\cite{Adhikari:2017esn,cosinebg}. 
This was confirmed by the observation of a 46.5~keV $\gamma$ peak that is a characteristic of $^{210}$Pb. 
In the $^{222}$Rn decay chain, $^{210}$Po (half-life $t_{1/2}$ = 138 days) is produced by the beta decay of $^{210}$Pb ($t_{1/2}$ = 22 years) and $^{210}$Bi ($t_{1/2}$ = 5 days), and subsequently decays via $\alpha$ emission to $^{206}$Pb. 
Thus, if the $^{210}$Po contamination is zero at the time the $^{222}$Rn exposure occurs, it will grow with a characteristic time of $\tau_{\rm ^{210}Po}$ = 200 days until an equilibrium level is reached. 
Thus, the total alpha rate over time can provide information about the amount of the $^{210}$Pb contamination. 
We modeled the measured alpha rate ($N$) as following formula, 
$$ N = N_{\rm Pb210}(1-e^{-(t-t_0)/\tau_{\rm ^{210}Po}})+C.$$
where $N_{\rm Pb210}$ is asymptotic $^{210}$Pb amount at infinite time, $t_0$ is contamination date fixed at the crystal was grown as summarized in Table \ref{crystal_growth}, and $C$ is the total alpha rate without $^{210}$Po decay. 
We can determine the $N$ from the mean time distribution~(Fig.~\ref{fig:alpha}).  
Figure~\ref{alphatime} shows the measured alpha rates in the crystals as a function of data-taking time. The time-dependent fit with the $^{210}$Po time-dependent model overlaid.   
From these fits, the $^{210}$Pb level in NaI-025 is 3.80$\pm$0.30~mBq/kg and, no significant $^{210}$Pb is observed for NaI-034 \& NaI-035 as one can see in Fig.~\ref{alphatime}. 
The measured level of $^{210}$Pb in NaI-036 (0.42$\pm$0.27 mBq/kg) is consistent with the $^{226}$Ra contamination (0.45$\pm$0.05 mBq/kg).

\subsubsection{$^{232}$Th background}
\begin{figure*}[htbp]	
  \centering 
	\begin{tabular}{ccc}
  \includegraphics[width = 0.33 \textwidth] {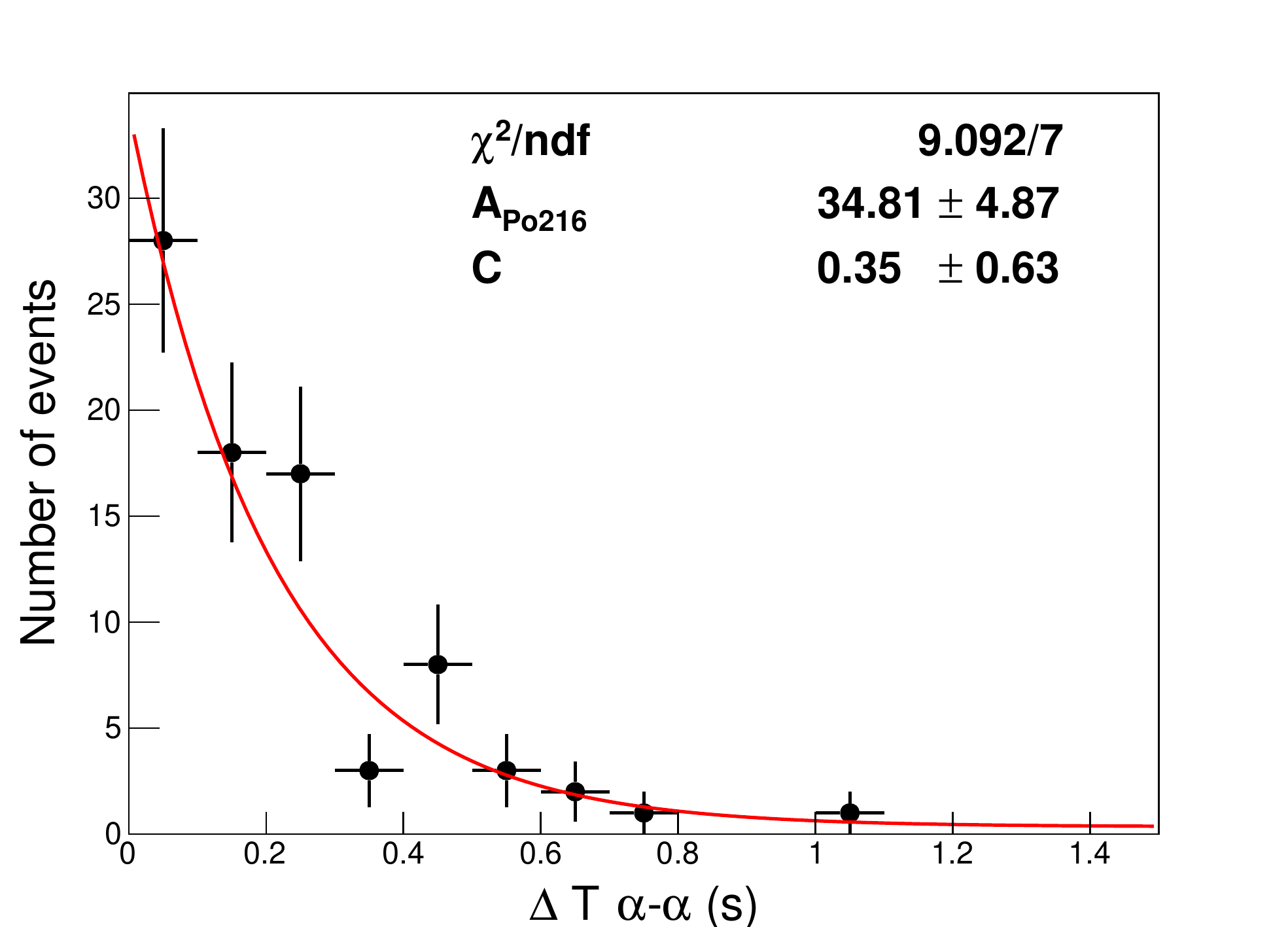} &
  \includegraphics[width = 0.33 \textwidth] {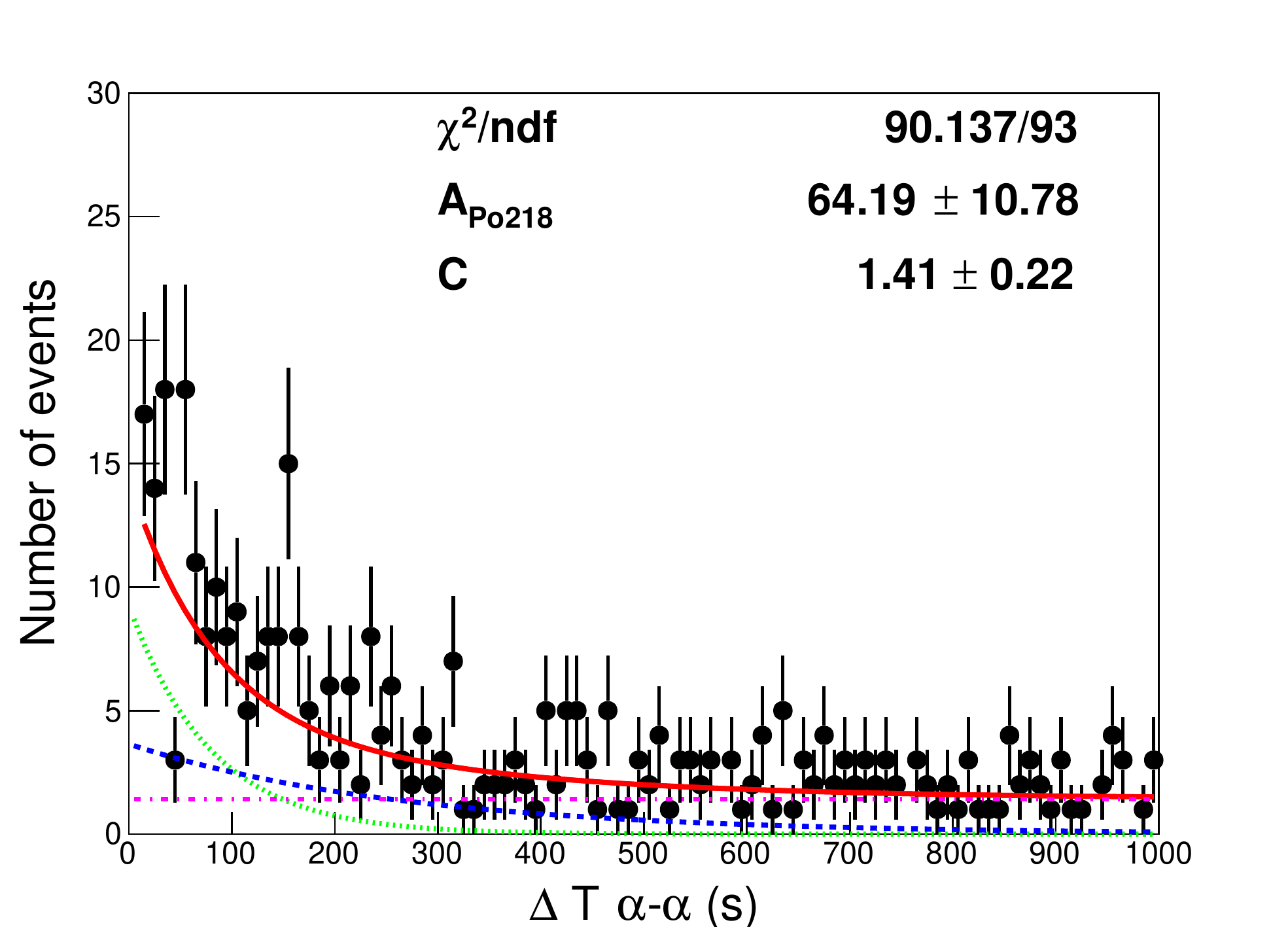} &
  \includegraphics[width = 0.33 \textwidth] {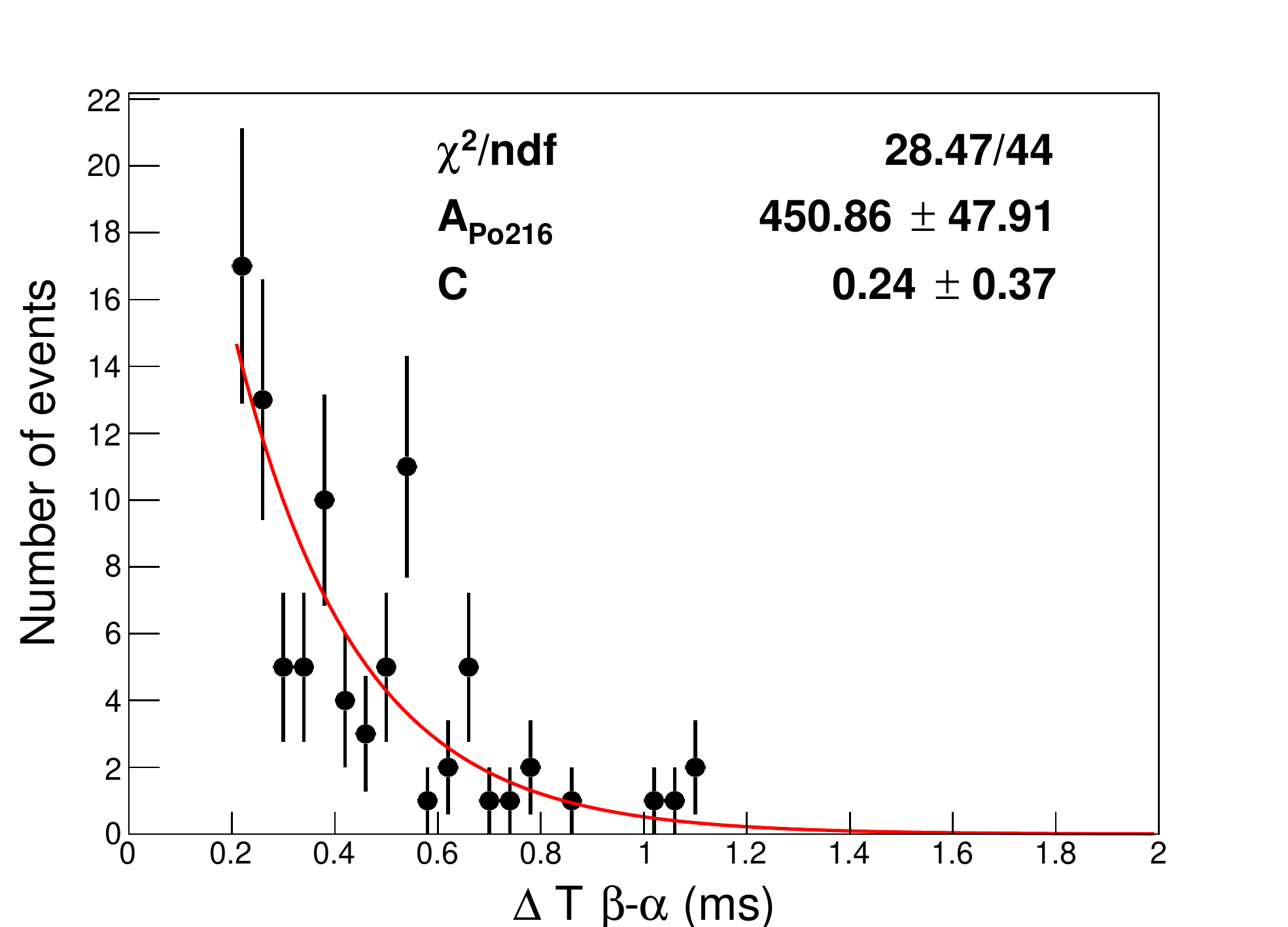} \\
	(a)  & (b)  & (c)  \\
	\end{tabular}
	\caption{The time difference~($\Delta$T) distribution of data~(points) and exponential fits~(red-solid line) between two successive $\alpha$-induced events are presented. Here decay time used in the fit is fixed to the known life-time of each radioisotope and measured rates are converted into activities ($A$) in unit of $\mu$Bq/kg. 
			$\Delta$T between two $\alpha$ decays of (a) $^{220}$Rn$\rightarrow ^{216}$Po~(half-life 0.145~s)  and (b) $^{224}$Ra$\rightarrow ^{220}$Rn~(green-dotted line, half-life 55.6 s), where the activity of $^{220}$Rn is constraint to $^{216}$Po measurement in (a), and $^{222}$Rn$\rightarrow ^{218}$Po~(blue-dashed line, half-life 186~s) together with random coincidence events~(purple-dashed-dotted line), in the NaI-034 crystal; (c) $\Delta$T between successive $\beta$ and $\alpha$ events from $^{214}$Bi$\rightarrow ^{214}$Po~(half-life 0.164~ms) in the NaI-036 crystal are shown. 
	Due to dead time of the data acquisition system, events with $\Delta$T less than 0.2~ms were not recorded. }
  \label{alpha_alpha}
\end{figure*}

Contaminants from the $^{232}$Th decay chain can be identified by studying three-alpha delayed time coincidences of  $^{224}$Ra, $^{220}$Rn and $^{216}$Po, where the half-lives are 3.66 day, 55.6 s, and 0.145 s, respectively. 
Due to the short half-life of $^{216}$Po ($t_{1/2}$=0.145~s), identification of two $\alpha$ particles from $^{220}$Rn and $^{216}$Po is relatively straight-forward and background free, as can be seen in Fig.~\ref{alpha_alpha}(a). With the selection of two consecutive $\alpha$ signals, $^{224}$Ra can also be identified. 
No significant  $^{232}$Th $\alpha$-decay events in NaI-025 and NaI-036 are measured. 
However, the presence of $\alpha$-$\alpha$ delayed coincidence events in NaI-034 and NaI-035 is evident in Fig.~\ref{alpha_alpha} with measured activity levels of 35$\pm$5~$\mu$ Bq/kg and 7$\pm$2~$\mu$ Bq/kg for NaI-034 and NaI-035, respectively. 
Assuming equilibrium of the $^{232}$Th chain, this is similar to the 2$-$31 $\mu$Bq/kg level in the DAMA crystals.

\subsubsection{$^{238}$U background}
$^{238}$U is one of the most common radioisotopes in nature primarily because of its long decay time. 
A study of its decay products allows us to understand its contamination level. 
Delayed coincidence $\alpha - \alpha$  events with $t_{1/2}$ = 3.10 min time difference 
from $^{222}$Rn$\rightarrow^{218}$Po can be used to infer the $^{238}$U contamination levels, as shown in Fig.~\ref{alpha_alpha}(b). 
The measured rate for $^{224}$Ra$\rightarrow^{220}$Rn ($t_{1/2}$ = 55.6 s) is extracted from the $^{232}$Th level, which is determined from $^{220}$Rn$\rightarrow^{216}$Po decays. 
In this way, we infer $^{218}$Po levels of 64$\pm$11 $\mu$Bq/kg and 8$\pm$6 $\mu$Bq/kg for NaI-034 and NaI-035, respectively. 
Because of the high total $\alpha$ rates in NaI-025 and NaI-036, we cannot measure 3~min half-life $\alpha - \alpha$ delayed coincidence events. 

Another method for studying the $^{238}$U chain exploits the 164.3~$\mu$s half-life $^{214}$Po $\alpha$-decay, which
follows its production via the $\beta$-decay of $^{214}$Bi.
Due to the 200~$\mu$s dead time of the trigger system, only events in the higher end of the exponentially falling distribution are acquired but the number of these events is sufficient to measure the $^{238}$U contamination (see in Fig.~\ref{alpha_alpha}(c)). 
The measured activity calculation includes a correction for the effect of the inefficiency caused by the 200~$\mu$s dead time. 
NaI-036 shows a relatively large $^{214}$Po contamination of 451$\pm$48 $\mu$Bq/kg, while NaI-025, NaI-034, and NaI-036 show contaminations of 26$\pm$7 $\mu$Bq/kg, 46$\pm$7 $\mu$Bq/kg, and 13$\pm$6 $\mu$Bq/kg, respectively. 
The contamination in NaI-036 is considerably larger than those of the DAMA crystals~(8.7$-$124 $\mu$Bq/kg) but the other crystal contamination levels are similar to those of DAMA. 

The energy spectrum of $\alpha$ events in Fig.~\ref{fig:alpha}(b) can be used to determine whether or not the $^{238}$U chain is in equilibrium, especially for NaI-036. The $\alpha$ energy spectrum of NaI-036 can be well explained by $^{226}$Ra decays with a 0.45$\pm$0.05 mBq/kg activity with no higher components from the $^{238}$U chain. At the start of the measurement time period, we did not see evidence for $^{210}$Po decay events in the $\alpha$ spectrum but the evolution plot in Fig.~\ref{alphatime}(d) can be explained by an approximately 0.45~mBq/kg of the $^{210}$Pb contamination coming from $^{226}$Ra. 
The $^{226}$Ra in NaI-036 can be attributed to impurities contained in the TlI powder.
Specially, the SA powder (initial loading 0.1 mol\% of NaI-034) and SA beads (initial loading 1 mol\% of NaI-036) contribute consistent contaminations of $^{226}$Ra to the crystals. 
If this is the case, we expect that the use of TlI beads from Alpha Aesar (NaI-035) could
reduce the $^{226}$Ra contamination significantly.

\subsection{External Background}
In the COSINE-100 experiment, the external background contributions in the region of interest~(ROI) are negligible~\cite{cosinebg} because of the active veto provided by the 2,200~L liquid scintillator~\cite{Adhikari:2020asl}. 
However, the R\&D setup contains a significant level of external background that is more than 10 times that for a full-size COSINE crystal~\cite{Adhikari:2017gbj}. 
Because the PMTs are the main contributor to the external background, a small crystal will have relatively larger background contribution per kilogram from external sources. Since we use the same type of PMTs in the COSINE-100 experiment, we can estimate the external background contribution using the GEANT4 simulation based on previous studies~\cite{cosinebg,Adhikari:2017gbj}. 

\subsection{Cosmogenic radionuclides}
\label{sec:cosmo}
The cosmogenic production of radioactive isotopes in the NaI(Tl) crystal contributes a non-negligible amount of background in ROI of the COSINE-100 experiment \cite{cosinebg,deSouza:2019hpk}. 
This is mainly due to long-lived nuclides such as $^{3}$H and $^{22}$Na.  
All crystals of the COSINE-100 experiment were produced at the Alpha Spectra company located at Grand Junction, Colorado, USA. 
Due to its high altitude of about 1,400~m above sea level as well as the long delivery time to Korea, the crystals were exposed to a significant flux of cosmic-ray muons and their spallation products and, as a result, we measure a relatively high level of cosmogenic radioactive isotopes. 
However, the crystals that are used for this study were all grown in Daejeon, Korea (70~m in altitude) and delivered underground in a short time period (typically within a few weeks). 
Based on our previous study, described in Ref.~\cite{deSouza:2019hpk}, one-month exposure time near sea level can produce 0.004~mBq/kg of $^{3}$H and 0.05~mBq/kg of $^{22}$Na. 
These activities are an order of magnitude lower than measured activities in COSINE-100 C6, which are 0.11~mBq/kg and 0.73~mBq/kg~\cite{deSouza:2019hpk} of $^{3}$H and $^{22}$Na, respectively. 
Thus, we expect that the background contributions from these long-lived cosmogenic radioisotopes will be sufficiently low for the COSINE-200 experiment.

\begin{table*} %[ht]
		\caption{Summary of the measured and fitted radioactive contaminants in the modeling of the NaI-035 crystal. }
\label{table:fits}
  \centering
\begin{tabular}{llllll}
  \hline                                                                                
 Background sources         & Isotopes & \multicolumn{3}{c}{Activities(mBq/kg)} \\ \cline{3-5}
	 && Measured &Input constraint& Fitted \\
	 \hline
	
	\multirow{4}{*}{Internal} & $^{238}$U&0.011$\pm$0.004 & $\pm$0.004 & 0.01$\pm$0.01    \\
         & $^{228}$Th & 0.007$\pm$0.002 &$\pm$0.002 & 0.006$\pm$0.004  \\
         & $^{40}$K   & 0.39$\pm$0.18  & $\pm$0.18& 0.52$\pm$0.18  \\
         & $^{210}$Pb & 0.014$\pm$0.022& $\pm$0.022& 0.009$\pm$0.009 \\
   \hline
  	\multirow{10}{*}{Cosmogenic}  &$^{125}$I&$-$ & $-$ & 0.39$\pm$0.01    \\
         & $^{121}$Te & $-$    &$-$& 0.80$\pm$0.03   \\
         & $^{121m}$Te & $-$     &$-$ & 0.06$\pm$0.01  \\
         & $^{123m}$Te  & $-$      &$-$& 0.10$\pm$0.09  \\
         & $^{125m}$Te  &  $-$      &$-$& 0.14$\pm$0.01   \\
         & $^{127m}$Te  &  $-$     &$-$& 0.19$\pm$0.02  \\
         & $^{109}$Cd  &  $-$     &$-$& 0.007$\pm$0.001  \\
         & $^{113}$Sn  &  $-$    &$-$ & 0.015$\pm$0.01  \\
         & $^{22}$Na & 0.05    &$\pm$0.03& 0.05$\pm$0.01  \\
         & $^{3}$H   & 0.004   &$\pm$0.002& 0.005$\pm$0.005  \\
 \hline
	\multirow{3}{*}{ NaI PMTs} & $^{238}$U &25$\pm$5   &$\pm$13& 27.2$\pm$11.2    \\
         & $^{232}$Th       &12$\pm$5&$\pm$6& 16.8$\pm$1.2  \\
         & $^{40}$K& 58$\pm$5  & $\pm$29     & 58.6$\pm$18.0  \\
	 \hline
	\multirow{3}{*}{ CsI PMTs} & $^{238}$U  &78$\pm$4  &$\pm$39.1 & 81.0$\pm$14.6   \\
         & $^{232}$Th&    26$\pm$4   &$\pm$13& 22.5$\pm$2.3  \\
         & $^{40}$K& 504$\pm$72 &   $\pm$252   & 395.7$\pm$7.2  \\
	 \hline
\end{tabular}
\end{table*}

\section{Background modeling and prospects of COSINE-200}
\begin{figure*}[htbp]	
	\centering
	\begin{tabular}{cc}
	\includegraphics[width=0.49 \textwidth] {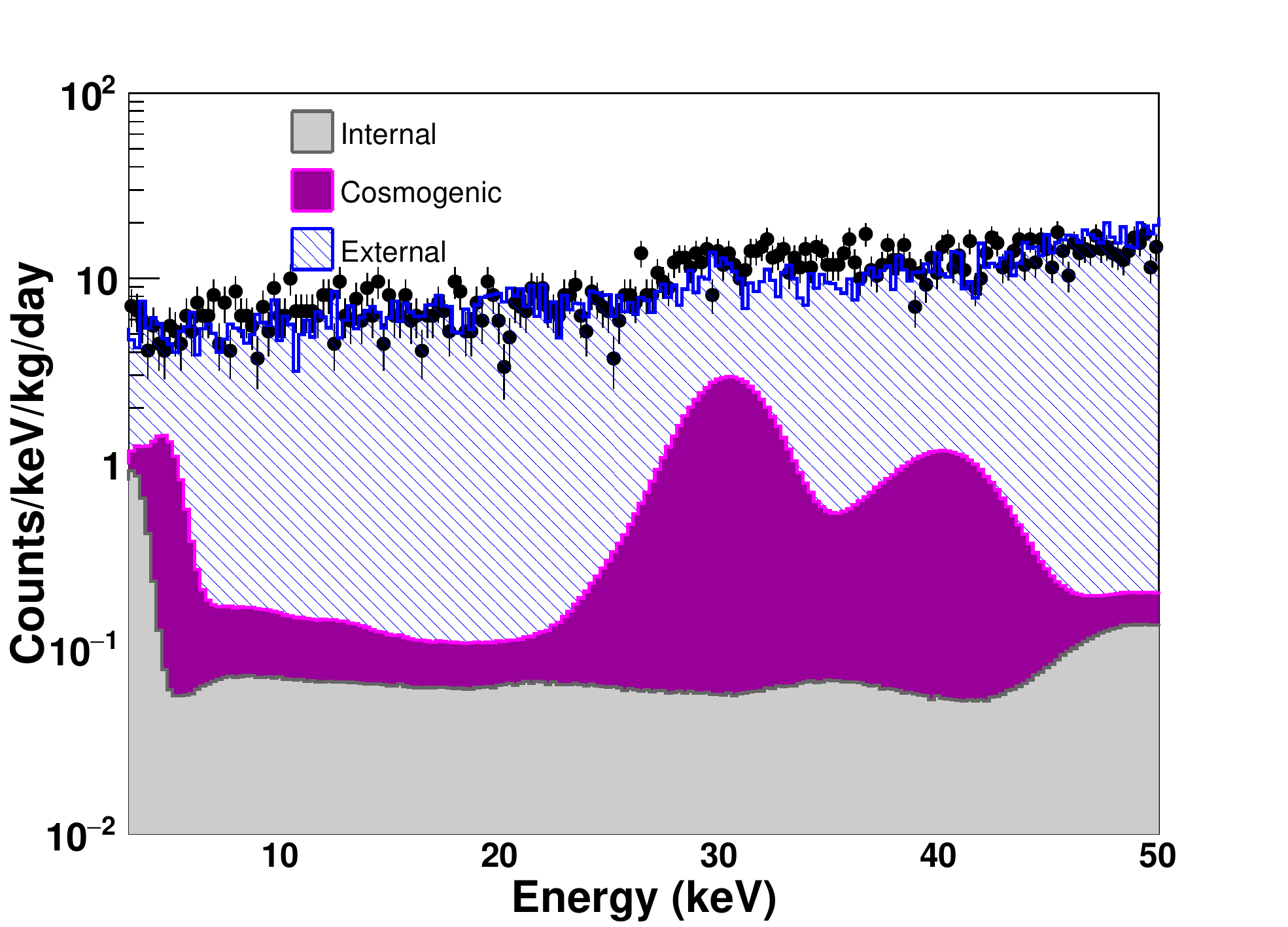} &
	\includegraphics[width=0.49 \textwidth] {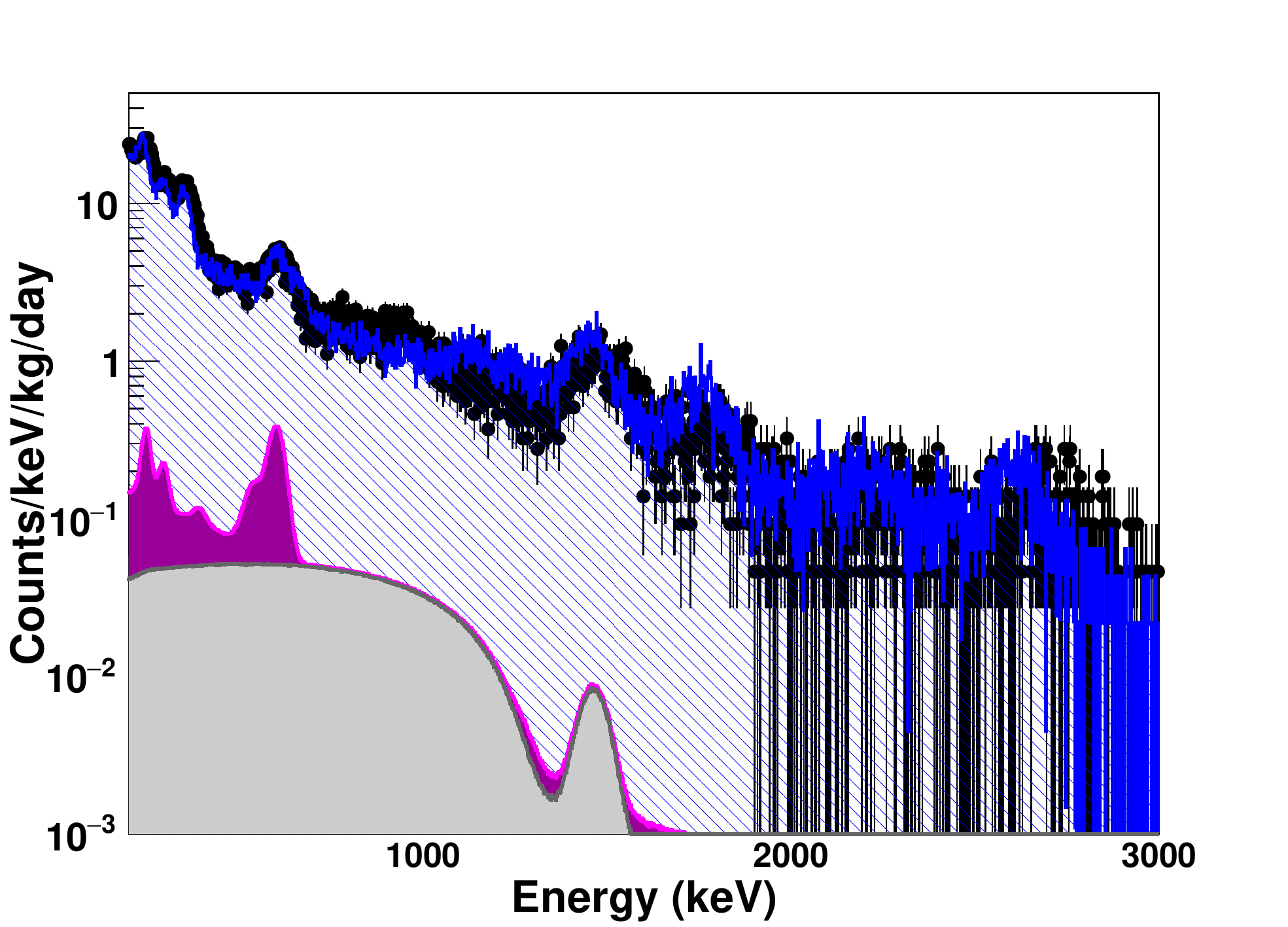} \\
	(a) Single-hit low-energy (3-60~keV) & (b) Single-hit high-energy (60-3000~keV) \\
	\includegraphics[width=0.49 \textwidth] {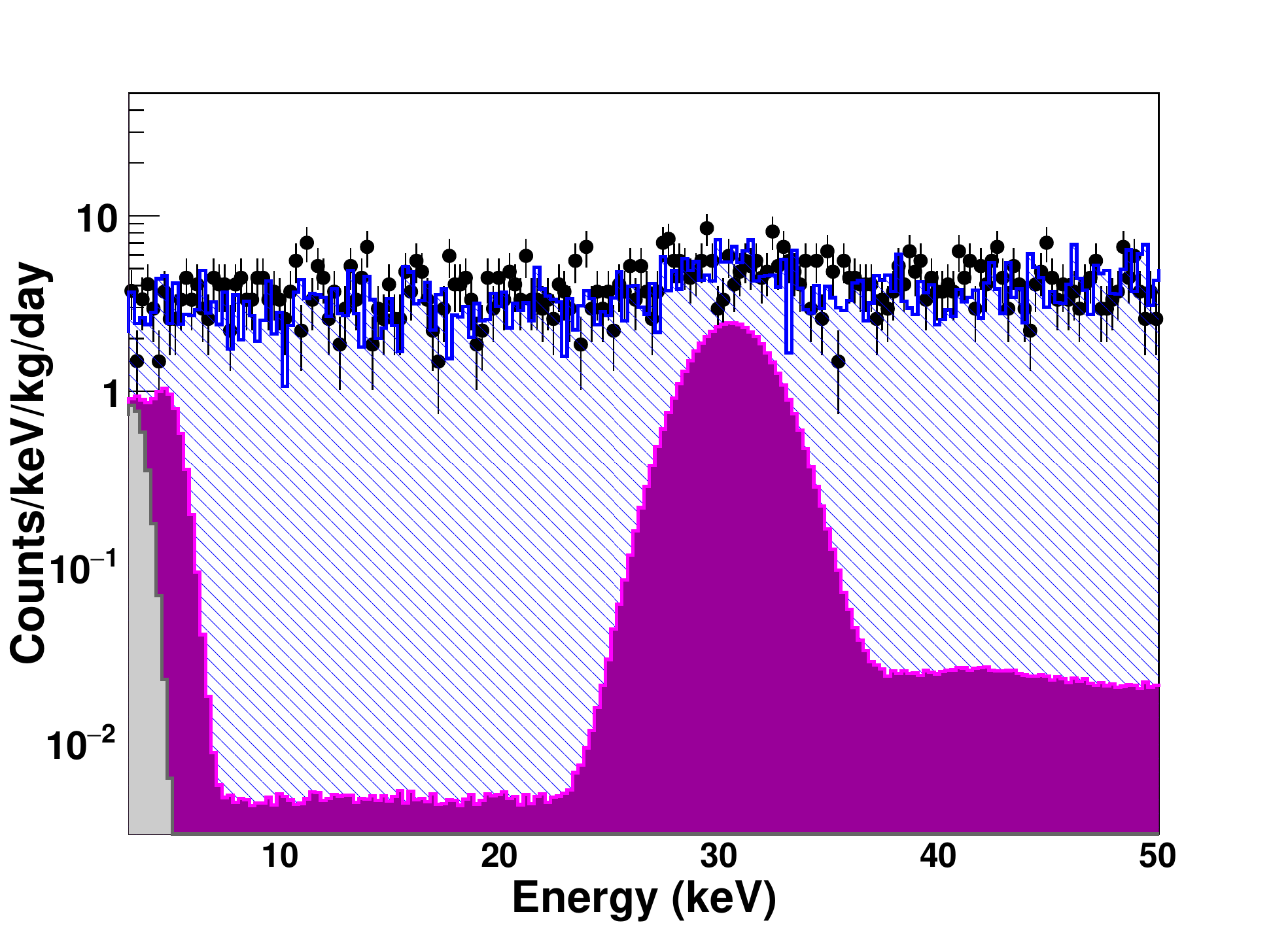} &
	\includegraphics[width=0.49 \textwidth] {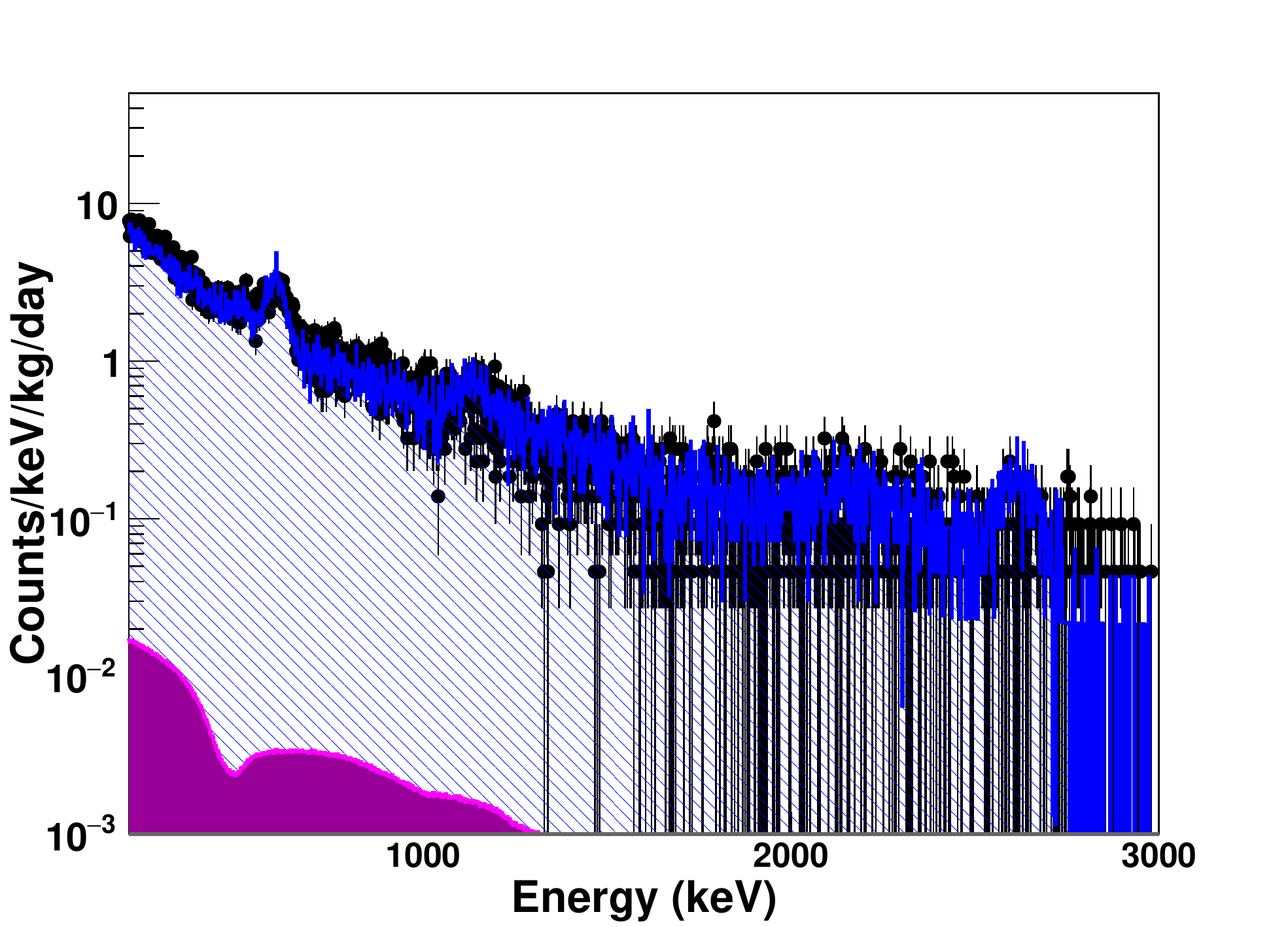} \\
	(c) Multiple-hit low-energy (3-60~keV) & (d) Multiple-hit high-energy (60-3000~keV) \\
	\end{tabular}
	\caption{Measured single-hit and multiple-hit background spectra of NaI-035 (black points) are determined for all simulated backgrounds using a simultaneous fit to the four-channels. The external component (blue-dashed filled) is the dominant contributor; this will be reduced with full-size detectors situated in the liquid scintillator veto system}
	\label{backgroundmodel}
\end{figure*}

For a quantitative understanding of background in these recently grown crystals, we use the GEANT4-based simulation that was developed for the background modeling of the COSINE-100 NaI(Tl) crystals~\cite{cosinebg,Adhikari:2017gbj}. 
We take values of the contamination levels discussed in Section~\ref{background} as input. 
Because of the low statistics of the underground $^{40}$K measurements, we use the ICP-MS results for $\rm ^{nat}$K from Table~\ref{crystal_growth}.

We use a log-likelihood method to fit the data. The fitting range is 3~keV$-$3~MeV and we perform a simultaneous fit of four-channels : single-hit low-energy (3$-$60~keV) and high energy (60~keV$-$3~MeV), multiple-hit low-energy and high energy. 
Here, multiple hit corresponds to events in which there is one or more coincident hits recorded in any of the surrounding CsI(Tl) crystals. 
The internal backgrounds are constrained to the measured uncertainties. 
We do not consider surface $^{210}$Pb components in the crystals because the crystal surface is polished in a nitrogen gas environment and we do not see any indication of increases in energy spectra in the low-energy region that are characteristic of surface $^{210}$Pb contamination \cite{Yu:2020ntl}. 
The backgrounds from the PMTs that are attached to the NaI(Tl) and CsI(Tl) crystals are measured with a high purity Germanium detector~\cite{cosinebg,Adhikari:2017gbj}. 
These values are constrained to be within 50\% of the measured results because exact locations of these radioisotopes are uncertain. 
The long-lived cosmogenic radioisotopes are constrained to be within 50\% of its calculated production values that are estimated in Section~\ref{sec:cosmo} while other short-lived cosmogenic components are floated. 
Additional external backgrounds from the CsI(Tl) crystals and copper surface are also included as free parameters but the final results are cross-checked with previous measurements made with the same setup~\cite{Adhikari:2017gbj}. 

Figure~\ref{backgroundmodel} shows the fitted NaI-035 results for all of the simulated background components plotted as lines with different colors. 
The overall energy spectra are well matched to the data for both single-hit and multi-hit events. 
The small sizes of the crystals as well as the non-existence of an active veto detector in the R\&D setup results in the external background being the dominant contributor to these spectra. 
However, this contribution will be significantly reduced with the large, COSINE-100-sized, crystals and the liquid scintillator veto system that will be used in the upcoming COSINE-200 setup. 
Table~\ref{table:fits} shows a summary of input values and fitted results from this background modeling. 

\begin{figure*}[htbp]	
	\centering
	\begin{tabular}{ccc}
	\includegraphics[width=0.33 \textwidth] {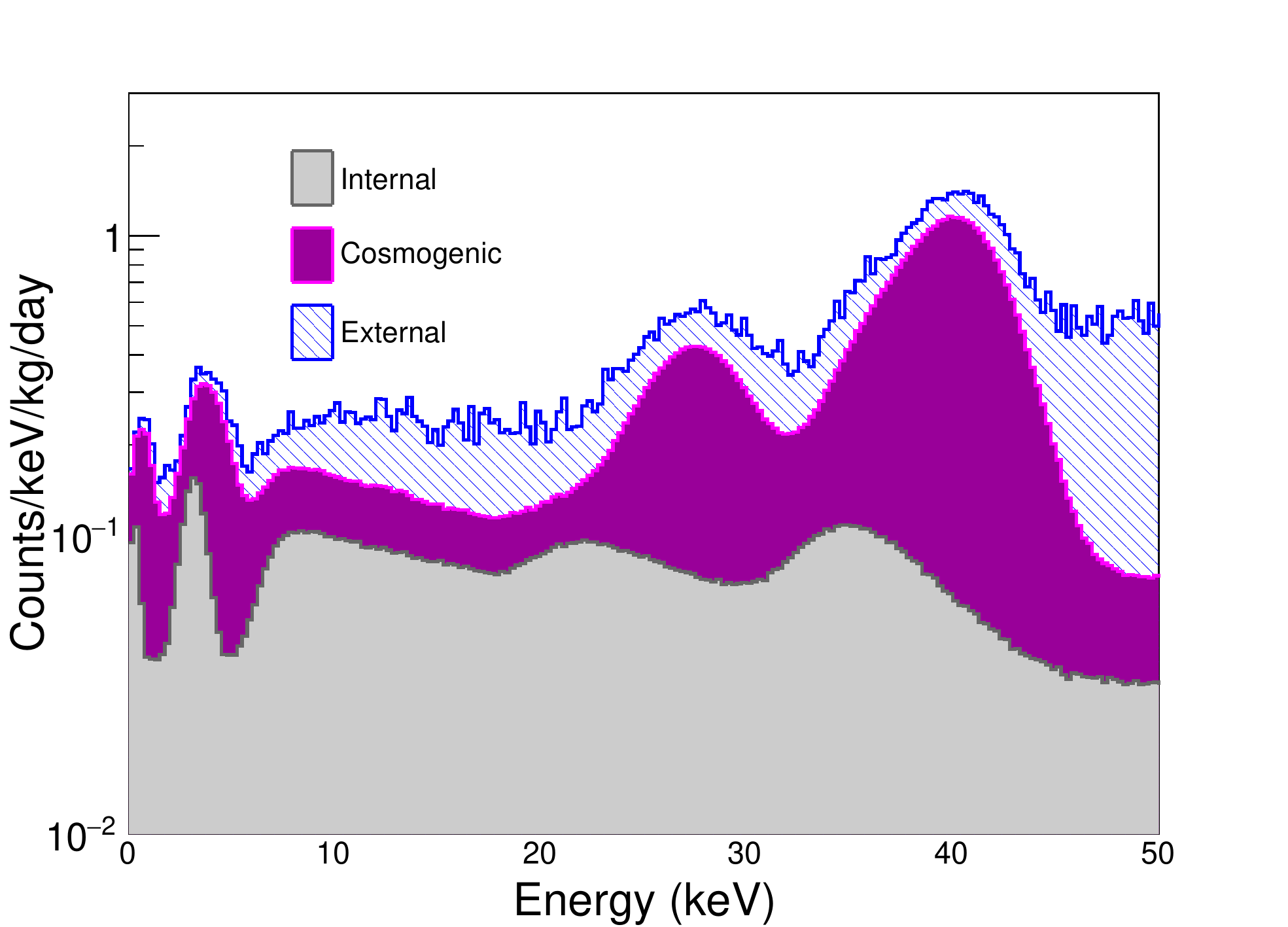} &
	\includegraphics[width=0.33 \textwidth] {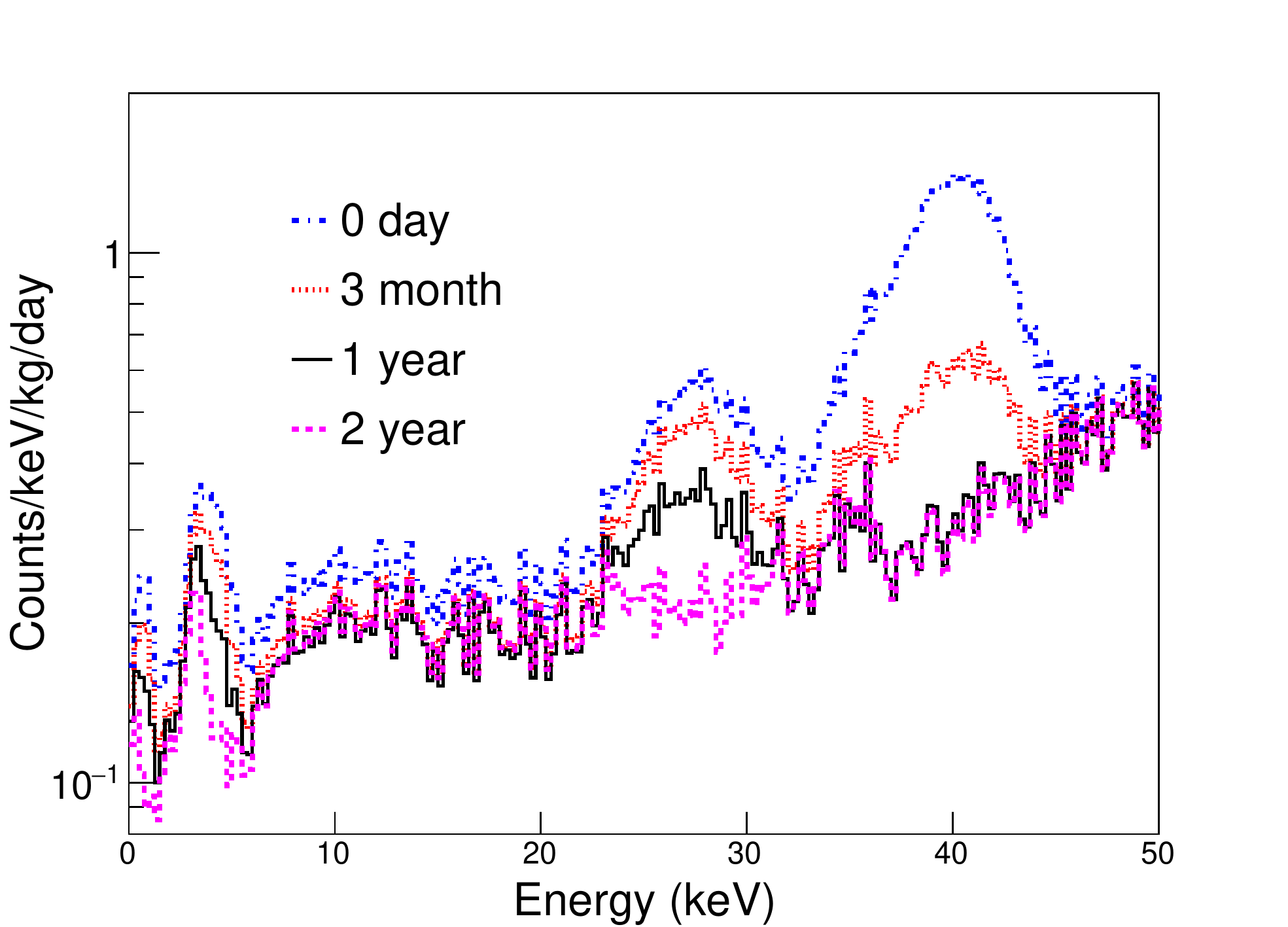} &
	\includegraphics[width=0.33 \textwidth] {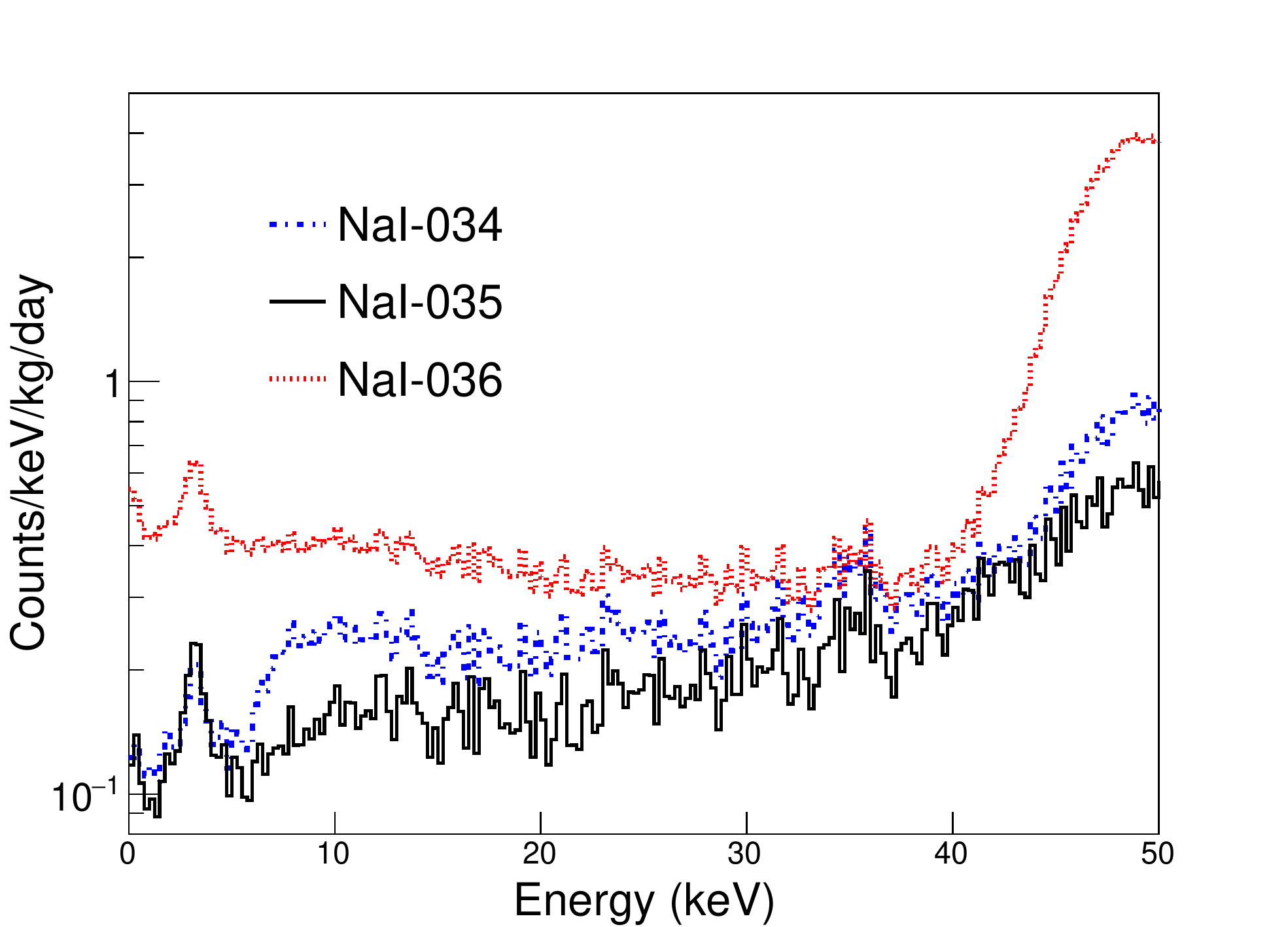} \\
	(a) & (b) & (c) \\
	\end{tabular}
	\caption{(a) The expected background spectrum of a COSINE-200 crystal for low-energy single-hit events assuming the same internal and cosmogenic radioisotopes as NaI-035 but, installed in the COSINE-100 shield with an enlarged crystal size (12.5 kg). (b) The time evolution of the background level due to decays of short-lived cosmogenic radioisotopes after 0 days (blue-dashed-dotted line), 3 months (red-dotted line), 1 year  (purple-dashed line) , and 2 years (black-solid line). (c) The expected background spectra of the COSINE-200 detectors assuming background levels obtained from NaI-034 (blue-dashed-dotted line), NaI-035 (black-solid line), and NaI-036 (red-dotted line) after being underground for two years. }
	\label{cosine200}
\end{figure*}

To estimate the expected background in the COSINE-200 crystal, we generate simulated background spectra assuming the same internal and cosmogenic radioisotopes in the crystals. 
Because of the relatively small contribution, the background modeling fit does not provide more accurate results for measurements of a few internal and cosmogenic $^{3}$H components, as shown in Table~\ref{table:fits}. The more accurate results between the measured activities shown in Table~\ref{table:crystals} and the fitted activities shown in Table~\ref{table:fits} are used.
We consider the crystal size enlarged to that of the C6 (12.5~kg) and installed inside a COSINE-100-like shield. 
The latter consists of a 2,200~L liquid scintillator tank, a 3 cm thick copper, a 20 cm thick lead, and an array of plastic counters for muon tagging~\cite{Prihtiadi:2017inr} as described in Refs.~\cite{Adhikari:2017esn,Ha:2018obm}.
We assume levels of external background that are similar to those seen in the COSINE-100 detector~\cite{cosinebg}.  
Because of significantly reduced $^{210}$Pb contaminations, the internal background contributions are reduced. 
We also expect that the lower exposure to cosmic muons will significantly reduce the level of long-lived cosmogenic such as $^{3}$H and $^{22}$Na. 
Figure~\ref{cosine200}(a) shows the expected background spectrum for the COSINE-200 detector assuming the internal background as NaI-035 background measurements in Table~\ref{table:fits}. 
By taking advantage of the 2,200~L active veto detector as well as the enlarged detector size, a significantly reduced background in low-energy single-hit events is expected. 
In the 1~keV to 6~keV ROI, the expected background level is approximately 0.33 counts/kg/day/keV, which is below that of the DAMA experiment. 
After time spent underground, the short-lived cosmogenic radioisotopes will decay away as shown in Fig.~\ref{cosine200}(b). 
After two years, the background is expected to decrease to about 0.13 counts/kg/day/keV. 
If we consider different background levels obtained from NaI-034 and NaI-036, the expected background from the COSINE-200 detectors would  be slightly increased as can be seen in Fig.~\ref{cosine200}(c).  However, it is still less than 1 counts/kg/day/keV that is target background level for the COSINE-200 experiment~\cite{adhikari16}. 

\section{Conclusion}
In the quest for dark matter direct detection, the origin of the DAMA modulation signal remains the subject of a long-standing debate. 
To expand the currently ongoing COSINE-100 experiment and reach a model-independent conclusion, we propose the COSINE-200 experiment with 200~kg ultra-pure NaI(Tl) crystals.
As a part of a program to reduce background levels, we have grown four small-size NaI(Tl) crystals in our laboratory. 
In this article, we present results from performance studies of these R\&D devices that demonstrate the feasibility of the production of low-background NaI(Tl) crystals suitable for the COSINE-200 experiment. 
It is necessary to have further R\&D on full-size crystals, and this is currently underway. 
Assuming the full-size crystals of similar quality as the NaI-035 detector that is installed in the COSINE-100-like shield,
our study shows that an expected background level will be less than 0.5 counts/kg/day/keV and  lower than those of the DAMA crystals. 
In this case COSINE-200 will be able to produce a definitive and unambiguous test of the DAMA experiment's annual modulation signal.

\begin{acknowledgements}
	We thank the Korea Hydro and Nuclear Power (KHNP) Company for providing underground laboratory space at Yangyang. This work is supported by the Institute for Basic Science (IBS) under project code IBS-R016-A1. 
\end{acknowledgements}

\bibliographystyle{JHEP}       

\providecommand{\href}[2]{#2}\begingroup\raggedright\endgroup
\end{document}